\documentclass[12pt,leqno,twoside]{article}
\usepackage{amsmath,amssymb,amsfonts,euscript,latexsym}
\usepackage[dvips]{graphics}
\bibliography{bb}
\def\md{\medskip}
\def\ds{\displaystyle}

\def\un{1\hskip-3pt\mathrm I}

\providecommand{\eqzero}{\setcounter{equation}{0}}

\begin{document}
\begin{titlepage}
\pagenumbering{arabic}
\begin{center}
{\bf{\large{Masslessness in $n$-dimensions}}}\footnote{To be published
in {\it Reviews in Mathematical Physics}, vol.10 (1998)}
\vskip2cm

{\sc{Eugenios}} ANGELOPOULOS and {\sc{Mourad}} LAOUES
\vskip1cm

{\footnotesize{

Laboratoire Gevrey de Math\'ematique Physique
\md

UNIVERSIT\'E DE BOURGOGNE
\md

9 avenue Alain Savary
\md

B.P. 400, F-21011 Dijon Cedex, France
\md

[E-mail: {\tt physmath@u-bourgogne.fr]}}

\vskip1cm

July 10, 1997}

\end{center}

\vskip2cm

{\sc{abstract}}.
We determine the representations of the ``conformal'' group
${\overline{SO}}_0(2, n)$, the restriction of which
on the ``Poincar\'e'' subgroup ${\overline{SO}}_0(1, n-1).T_n$
are unitary irreducible. We study their restrictions
to the ``De Sitter'' subgroups ${\overline{SO}}_0(1, n)$
and ${\overline{SO}}_0(2, n-1)$ (they remain irreducible or decompose
into a sum of two) and the contraction of the latter to
``Poincar\'e''. Then we discuss the notion of masslessness
in $n$ dimensions and compare the situation for general $n$
with the well-known case of 4-dimensional space-time, showing the
specificity of the latter.

\end{titlepage}
\newpage



\section*{Introduction}

The formulation of a unifying theory which would include all fundamental
interactions of physics is still an open problem, though the need for it
was realized early in this century, when modern physics (relativistic
and quantum) was introduced and in spite of unsuccessful efforts of
many of its founders. A major difficulty consists in unifying the so-called
gauge interactions and gravitation. A number of approaches to this question,
appearing in various models such as the Kaluza-Klein theories or
supergravity \cite{Fro,SaSth}, use an imbedding of the four-dimensional
Minkowski space-time into a higher dimensional one (that is, $\mathbb R^n$
endowed with a $(1, n-1)$-Lorentz metric), then getting rid of the redundant
spatial dimensions by various techniques, such as spontaneous compactification.

\md

On the other hand, an essential feature of relativity is the boundary
character of the speed of light, which implies qualitatively distinct
behaviours for massless and for massive particles.
Mathematically, this is expressed by the (1,3) signature of Minkowski
space and the distinction between the massless and massive case is
kinematically expressed by distinct types of unitary irreducible
representations (UIR) \cite{Win} of the kinematic group, the Poincar\'e
group $\mathcal P$.

\md

Masslessness in four dimensions has been quite well studied from the
group theoretical point of view. We shall start by recalling the relations
between $\mathcal P$ and the De Sitter groups.
Let $M_\rho$ be a four-dimensional manifold with constant curvature $\rho$.
Its isometry group is the De Sitter group $G_\rho$, which is
isomorphic to $SO_0(2,3)$ (resp. $SO_0(1,4)$) if $\rho > 0$
(resp. $\rho< 0$); a physical reason for the introduction of the curvature
is that it provides efficient invariant infrared regularization
in the limit of zero curvature \cite{FlFr}. $M_\rho$ is isomorphic to the
homogeneous space $G_\rho / {\mathcal L}$ where $\mathcal L$ is the Lorentz
group $SO_0 (1,3)$.

\md

In the limit $\rho = 0$, $M_\rho$ becomes the (flat) Minkowski space and
$G_\rho$ contracts to the Poincar\'e group $\mathcal P$. Concerning
representations, it may however happen that two nonequivalent UIR of $G_\rho$
contract to the same massless representation of $\mathcal P$. Moreover,
if $\rho < 0$, the representations of $\mathcal P$ one gets by contraction have
an unbounded energy spectrum.

\md

Now the conformal group $G = SO_0(2,4)$ acts on compactified Minkowski space.
Massless UIR of $\mathcal P$ with discrete helicity extend {\it uniquely}
to UIR of $G$ acting on the same Hilbert space, and are the only
ones with this property, besides the trivial \cite{AF}. It turns out that
if such a UIR is extended to $G$, then restricted to the De Sitter subgroup
$SO_0(2,3)$, and finally contracted to $\mathcal P$, the initial representation
of $\mathcal P$ is recovered.
Therefore, from a kinematical point of view, the representations of the
De Sitter group $SO_0(2,3)$ thus obtained provide a satisfactory tool for
the extension of masslessness on $M_\rho$. Furthermore one can define a gauge
theory in the sense of Gupta-Bleuler and show that massless particles
propagate on the light cone, and so on \cite{AFFS,Tod}.

\md

Since the geometry of the $n$-dimensional Minkowski space-time is determined
by its Lorentzian metric $(1, n-1)$, its kinematic group is ${\mathcal P}_n =
SO_0(1, n-1). T_n$, and all groups related to it (Lorentz, conformal,
De Sitter) are similarly defined. It is then natural to ask which properties
of massless UIR of the Poincar\'e group extend to $n$ dimensions,
independently of other considerations: this is the purpose of this paper.

\md

More precisely, we shall study here the following topics:

\smallskip

i) Which UIR $U$ of ${\mathcal P}_n$ extend to irreducible representations
$d$ of the corresponding conformal group $G_n = SO_0(2, n)$?
To be precise we are dealing with projective representations but it turns out
that all the interesting ones are representations of the twofold covering
for all groups concerned.\\
ii) Is the extended representation $d$ unitary, and is it unique?\\
iii) Is the restriction $d'$ of $d$ on the De Sitter subgroups $SO_0(1, n)$
and $SO_0(2, n-1)$ irreducible?\\
iv) Can $d'$ be contracted to the initial one, $U$?

\md

The answers to these questions for $n=4$ were given in \cite{AFFS}
(except for those concerning $SO_0(1, 4)$), some of the results being anterior
to that paper: only (and all) UIR with zero mass and discrete helicity
do extend to $G_n$ with uniqueness and unitarity; the restriction to
$SO_0 (2, 3)$ is irreducible unless the inducing representation of the
little group is trivial (zero helicity) and it can be contracted back
to the initial UIR.

\md

For structural reasons (all groups concerned have similar structure and real
rank at most 2), a straightforward generalization was expected, at least for
$n$ even. It turned out, however, that though most features do indeed
generalize, the constraints on the existence of the extension increase
significantly with $n$. To be more precise, $U$ must again be massless,
that is induced by a UIR $S$ of the little group $SO_0(n-2).T_{n-2}$
(the Euclidean group in $n-2$ dimensions).
Not only $S$ has to be trivial on the translations (the analogue of discrete
helicity), but it must also be a {\it very degenerate} representation of
$SO_0(n-2)$.

\md

Acceptable $S$ are characterized by a discrete parameter $2s \in \mathbb Z$
when $n$ is even ($2s$ is the helicity for $n=4$), while for $n$ odd $S$
must be either trivial or spinorial. Also, the results for the De Sitter
subgroup generalize, with the sole exception of the irreducibility of $d'$
for odd $n$: it reduces into the direct sum of two simple factors for
spinorial $S$ too. As far as $SO_0(1, n)$ is concerned, $d'$ is always
irreducible.

\md

We therefore see once more, in this simple (kinematical) group theoretical
study, that the 4-dimensional space-time of special relativity and the
related universes with constant curvature are really special. In higher
dimensions the notion of masslessness becomes more involved and
requires, in addition to zero mass, a degeneracy far greater than the
requirement of discrete helicity in 4 dimensions.

\md
The paper is organized as follows.
In Section 1 we fix notations for $G_n, {\mathcal P}_n$ and the normalizer
${\mathcal W}_n$ of ${\mathcal P}_n$ in $G_n$. Since we are interested in
projective UIR we also present their universal coverings; in fact, as we shall
see later, only twofold coverings are needed, corresponding to the
covering of $SO(n)$ by $Spin(n)$. We also identify the compactified
$n-$Minkowski space with the quotient ${G}_n / {\mathcal W}_n$
and describe the action of ${G}_n$ on it. We then establish the unitary dual
of ${\overline{\mathcal P}}_n$, using the orbit-stabilizer method, and discuss
the possibility of extending a UIR $U$ to ${\mathcal W}_n$.
Propositions 1.1 and 1.2 give the (expected) result:
$U$ must be massless, and induced by a representation $S$
with trivial restriction on the Euclidean translation subgroup.

\md

Section 2 is devoted to determine which among the representations $d$ of
${G}_n$ can be viewed as extensions of massless UIR of ${\mathcal P}_n$,
using Lie algebraic methods. We begin by expressing the
weight representations of the complexified $\mathfrak{so}(N)^{\mathbb C}$,
in a way which can be used both for $\mathfrak{so}(2, n)$ and for the compact
real form $\mathfrak{so}(n-2)$ of the little group. We next translate into
enveloping algebra properties the fact that the squared $n$-mass operator
$P^\mu P_\mu$ is mapped to $0$ by $d$, calling such a $d$ a massless
representation. After the study of low $N$, we determine the finite-dimensional
ones, parameterizing them by a discrete parameter (Thm 2.3).

We next study infinite-dimensional ones, showing that on every
$\mathfrak{so}(n)$-type the character of the $\mathfrak{so}(2)$
 which commutes with $\mathfrak{so}(n)$ is fixed and increases
 in absolute value with the
Casimir of $\mathfrak{so}(n)$, keeping a fixed sign. Moreover, the lowest
$\mathfrak{k}$-type must be a massless representation of $\mathfrak{so}(n)$
itself: this constraint has no effect when $n=4$, but does cut off a huge part
in general (Thm 2.4). Massless representations are unitary and possess an
extremal weight.

In the following paragraph we identify the UIR ${U}^S$ of
${\overline{\mathcal P}}_n$ which extend to massless UIR of the conformal
group ${\overline{G}}_n$.
The inducing $S$ must be a finite-dimensional massless UIR of $SO(n-2)$.
The expression of the generators of ${G}_n$ as differential operators is
uniquely determined by those of ${\mathcal P}_n$.

\md

Section 3 discusses De Sitter subgroups. Irreducibility of the restriction is
examined on $\mathfrak{k}$-types, which all have multiplicity one. As for the
contraction to Poincar\'e, the proof of \cite{AFFS} extends easily to the
general case. The paper ends with a few remarks, where in particular we
briefly recall and present in the light of the present study the known results
for the lower dimensional cases $n=3$ and $n=2$.

\hskip2cm

 {\bf{ACKNOWLEDGEMENTS}}

\md

The authors wish to thank Mosh\'e Flato for suggesting the problem and
his constant interest in this work and Daniel Sternheimer for helpful
criticism of the manuscript. This work was partially supported by
E.U. Program ERBCHRXCT940701.

\newpage

\section{Poincar\'e and Conformal Group in $n$-dimensions}\eqzero

\hskip1cm

 {\bf{a) The $n$-Poincar\'e group ${\mathcal P}_n$}}

Let $n\geq 3$ be a fixed integer. Let $\{e_\mu\}_{\mu \in J}$,
with $J = \{0, 1, \ldots, n-1\}$
be a basis of $\mathbb R^n$; let $J'= \{1, \ldots, n-1\}$
and define a quadratic form $q$,
such that:

\begin{equation}
q(e_0) = 1 = -q(e_{\mu'}) \hskip.5cm \forall \mu ' \in J'.
\end{equation}

The associated symmetric bilinear form is denoted by $g$,
with $g_{\mu \nu} = g (e_\mu, e_\nu)$, which is equal
to $q(e_\mu)$ if $\mu = \nu$ and equal to zero otherwise.
The quadratic space $(\mathbb R^n, q)$
will be denoted $\mathbb R^{1, n-1}$,
and called the {\it{n-Minkowski space}}. It can be
identified with its dual, the dual basis
being $\{e^\mu\}_{\mu \in J}$, with

\begin{equation}
e^0 = e_0 \hbox{\ and \ } e^{\mu '} = -e_{\mu'}\hbox{ for } \mu ' \in J'
\end{equation}

\noindent and we shall write $g^{\mu \nu} = g(e^\mu, e^\nu)$,
with the same properties as
$g_{\mu \nu}$. Any element $x\in \mathbb R^{1, n-1}$ has the form:

\begin{equation}
x = x_\mu e^\mu = x^\mu e_\mu, \hbox{ with } x_\mu =
g_{\mu \nu} x^\nu \hskip.5cm g_{\mu \nu} g^{\mu \lambda}
= \delta^\lambda_\mu
\end{equation}
\noindent where $\delta$ is the Kronecker symbol.

\medskip

 {\it{Remark:}} The Einstein summation convention over the set $J$
is used in (1.3). It will be used throughout this paper. The range of
summation will not always be the same, and we shall use distinct
index variables for distinct ranges of summation. For instance we shall write
$x_{\mu '} x^{\mu '}$ instead of
$\ds  -\hskip-.5cm\sum_{1 \leq \mu ' \leq n-1} (x_{\mu'})^2$,
using the primed letter $\mu '$ instead of $\mu$ to avoid confusion. If the
range of summation is $J$, greek letters $\lambda, \mu, \nu, \ldots$
will always be used.

\md

The connected component of the Lie group of linear transformations of
$\mathbb R^n$ which leave $g$ invariant, $SO_0 (1, n-1)$, will be called
the {\it{$n$-Lorentz group}} and denoted by ${\mathcal L}_n$.
Its maximal compact subgroup is $SO(n-1)$. The twofold covering of
the latter (universal
if $n>3$) will be denoted by $Spin (n-1)$ and the corresponding
covering of ${\mathcal L}_n$ by ${\overline{\mathcal L}}_n$.

\md

The abelian group of translations of $\mathbb R^k$,
homeomorphic to $\mathbb R^k$, will be denoted by $T_k$, for
$k \in \mathbb N$. The semidirect product ${\mathcal L}_n \cdot T_n$,
where ${\mathcal L}_n$
acts canonically on $T_n$, will be called the {\it{$n$-Poincar\'e group}}
and denoted by ${\mathcal P}_n$. Its twofold covering
(universal if $n > 3$) ${\overline{\mathcal L}}_n \cdot T_n$ will be denoted by
${\overline{\mathcal P}}_n$.

\md

The Lie algebra ${\mathfrak{p}}_n$ of ${\mathcal P}_n$ is spanned
by generators $X_{\mu \nu} = - X_{\nu\mu} \in {\mathfrak{l}}_n =$
Lie (${\mathcal L}_n$) and $P_\mu \in {\mathfrak{t}}_n =$ Lie ($T_n$),
with $\mu, \nu \in J$, satisfying the commutation relations

\bigskip

\noindent  (1.4a)\hskip-31pt \centerline{
$[X_{\lambda\mu}, X_{\nu \rho}] = g_{\mu \nu} X_{\lambda \rho}-
g_{\lambda  \nu} X_{\mu  \rho} - g_{\mu  \rho}
X_{\lambda  \nu} + g_{\lambda  \rho} X_{\mu  \nu}$}

\bigskip

 \noindent (1.4b)\hskip-31pt \centerline{
$[X_{\lambda \mu}, P_ \nu] = g_{\mu  \nu}
P_\lambda - g_{\lambda  \nu} P_\mu$}

\bigskip

 \noindent (1.4c)\hskip-57pt \centerline{$[P_{\lambda}, P_\mu] = 0$}

\bigskip

\addtocounter{equation}{+1}

The element $P_\mu P^\mu = g^{\mu  \nu} P_\mu P_ \nu$ of
the enveloping algebra ${\mathcal U}(\mathfrak{p}_n)$ commutes with
all generators. In the classical case $n=4$, when used in
theoretical physics, it gives the squared mass of a particle.

\vskip1cm

 {\bf{b) The $n$-conformal group $G_n$}}

Let $I=\{-1, 0, 1, \ldots, n\} =
J \cup {\hat{J}}$ with ${\hat{J}} = \{-1, n\}$.
Extend the basis $\{e_\mu\}_{\mu \in J}$ of $\mathbb R^n$
to the basis $\{e_A\}_{A\in I}$ of $\mathbb R^{n+2}$, and extend the
quadratic form $q$ to $\mathbb R^{n+2}$ by putting
$q(e_{-1}) = 1 = -q(e_n)$. The quadratic space thus obtained will be
denoted $\mathbb R^{2, n}$; the associated symmetric
bilinear form will be again denoted by $g$, with
$g_{AB} = g(e_A, e_B), g^{AB} = g(e^A, e^B)$ for the dual basis,
and $g_{AB} g^{BC} = \delta^C_A$.

\md

The connected group $SO_0(2, n)$ which conserves the bilinear
form $g$ will be called the {\it{$n$-conformal group}} and
denoted by $G_n$. Its maximal compact subgroup is
$SO(2) \times SO(n)$, with universal covering
(for $n\geq 3$) $\mathbb R \times$ $Spin(n)$, that is,
infinite-fold times twofold. The universal covering
of $G_n$ will be denoted ${\overline{G}}_n$.

\md

The Lie algebra ${\mathfrak{g}}_n$ of $G_n$ is spanned
by generators $X_{AB} = -X_{BA} (\hbox{\footnotesize {A, B}} \in I)$, with
commutation relations:

\begin{equation}
[X_{AB}, X_{CD}] = g_{BC} X_{AD} -
g_{AC} X_{BD} - g_{BD} X_{AC} + g_{AD} X_{BC}.
\end{equation}

We shall denote by $C$ the Casimir element of the enveloping
algebra ${\mathcal U} ({\mathfrak{g}}_n)$, defined by:

\begin{equation}
C={\frac{1}{2}} X_{AB} X^{BA} = {\frac{1}{2}}
X_{AB} g^{BC} X_{CD} g^{DA}.
\end{equation}

The stabilizer of the basis elements $e_{-1}, e_n$ is obviously
${\mathcal L}_n$. Moreover, the set of generators $X_{\mu, -1} \pm
X_{\mu, n} (\mu \in J)$ spans an $n$-dimensional
abelian subalgebra isomorphic to ${\mathfrak{t}}_n$, on which
${\mathfrak{l}}_n$ acts like (1.4b),
for either choice of the $\pm$ sign. The corresponding group
elements have the $(n+2)\times(n+2)$ matrix form:

\begin{equation}
\hbox{exp}t^\mu (X_{\mu,-1}\pm X_{\mu, n}) =\left[
\begin{array}{lcl}
1-q(t)/2\hskip.2cm & \hskip.2cm t \hskip.2cm &\hskip.2cm \pm q(t)/2\cr
  &   &  \cr
-t^{\#}\hskip.2cm & \hskip.2cm \un_n\hskip.2cm &\hskip.2cm \pm t^{\#}\cr
   &   &  \cr
\mp q (t)/2\hskip.2cm & \hskip.2cm\pm t\hskip.2cm & \hskip.2cm 1 + q(t)/2\cr
\end{array}
\right]
\end{equation}

\noindent where $t^{\#}$ is the column vector $(t_\mu)$ and $t$
the line vector $(t^\mu)$.

The two Poincar\'e subgroups thus obtained are
conjugated in $G_n$, through the involutionary mapping:
$$\Theta = {\mathrm {Ad}}_{\hbox{exp} (\pi X_{n-1,n})}.$$
\noindent We shall write hereafter

\begin{equation}
P_\mu = X_{\mu, -1} + X_{\mu, n}\hskip.5cm
; \hskip.5cm \hat{P}_\mu = X_{\mu, -1} -
X_{\mu, n}
\end{equation}

\noindent and we shall identify $T_n$ as the subgroup spanned
by exp$(t^\mu P_\mu)$; the ``other'' translation subgroup
will be denoted $\hat{T}_n$, and the corresponding $n$-Poincar\'e subgroups
${\mathcal P}_n$ and $\hat{\mathcal P}_n$. The twofold covering
${\overline{\mathcal P}}_n$ is a subgroup of
${\overline{G}}_n$ (for $n=3$ the universal covering of ${\mathcal P}_n$ is
not contained in ${\overline{G}}_n$).

The remaining generator $D = X_{n,-1}$ will be called the dilatation,
it commutes with the Lorentz generators and its nonzero commutation
relations are

\begin{equation}
[D, P_\mu] = P_\mu \hskip.5cm; \hskip.5cm [D,{\hat{P}}_\mu]
= -{\hat{P}}_\mu
\end{equation}

One also has:

\begin{equation}
[P_\mu, {\hat{P}}_\nu] = -2 (X_{\mu \nu} + g_{\mu \nu} D)
\end{equation}

The normalizer ${\mathcal W}_n$ of ${\mathcal P}_n$ in
$G_n$ is the semidirect product $Y_n \cdot T_n$, with
$Y_n = A \times (W\cdot {\mathcal L}_n)$, where
$A =\{\exp tD\}_{t \in \mathbb R}$ and
$W = \{1,w\}$ is a group of order two, with
$w= \hbox{exp}(\pi (X_{0,-1} + X_{n-1,n})).$
The action of $w$ on $P_\mu$ is given by:

\begin{equation}
{\mathrm Ad}_w (P_\mu) = \epsilon_\mu P_\mu;\hskip.2cm
 \epsilon_0 = \epsilon_{n-1} =
1;\hskip.2cm \epsilon_j =-1 \hbox{ if } 1\leq j \leq n-2.
\end{equation}

$W\cdot {\mathcal L}_n$ is the non-connected group $SO (1, n-1)$.
The group $ {\mathcal W}_n$ is a maximal parabolic subgroup of $G_n$.
The same holds for the normalizer
${\hat{\mathcal W}}_n = Y_n \cdot {\hat{T}}_n$ of ${\hat{\mathcal P}}_n$,
and one has the Bruhat-type decomposition:

\begin{equation}
G_n \simeq {\hat{T}}_n Y_n T_n =
{\hat{\mathcal W}}_n T_n = {\hat{T}}_n  {\mathcal W}_n
\end{equation}

\noindent that is, the set of elements which can be written in this form is a
Zarisky open in $G_n.$ To be more precise, ${\mathcal P}_n$
(resp. ${\hat{\mathcal P}}_n$) stabilizes the point $e=e_{-1} + e_n$
 (resp. $\hat{e} = e_{-1} - e_n$) of $\mathbb R^{n+2}$. The
orbit of $e$ under $G_n$ is the isotropic
cone minus the origin, that is $G_n/{\mathcal P}_n = Q =\{y, y
\in \mathbb R^{n+2} / y \neq 0 \hbox{ and } y_A y^A = 0\}$.
The group $A\times W$ sends $e$ to $\lambda e$ (and $\hat{e}$ to
$\lambda {\hat{e}}$), $\lambda \in \mathbb R -\{0\}$, so that
$G_n/ {\mathcal W}_n = {\mathcal C}_0 = Q/(\mathbb R - \{0\}) \cong (S_1
\times S_{n-1}) / \mathbb Z_2$
is the set of directions of $Q$. The translation group $T_n$ stabilizes
the direction $\lambda e$ and acts transitively on the complementary
subset of ${\mathcal C}_0$. Thus the complementary subset of
${\hat{\mathcal W}}_n T_n$ in $G_n$ is $\{g; ge \in e^{\bot}\}$.

\md

${\mathcal C}_0$ is thus diffeomorphic to the compactified $T^c_n$
of $T_n$, that is $T^c_n = \mathbb R^{1, n-1}
\cup {\mathcal C}_\infty$ where ${\mathcal C}_\infty$ is the
$(n-1)$-dimensional compactified ``light cone at infinity''
\cite{CGT}\cite{KuWo}\cite{Tod}. Writing $\mathbb R^{2, n}$ and
$\mathbb R^{1, n-1}$ as line vectors we have the imbedding
$\varphi$ from $\mathbb R^{1, n-1}$ to $Q$:

\begin{equation}
\varphi(t) = e' \hbox{exp} (t^\mu P_\mu)
\end{equation}

\noindent with $e'=(1, 0, \ldots, 0, 1)$; using (1.7) one has:

\begin{equation}
\varphi(t) = (1 -q(t), 2t, 1+ q(t)).
\end{equation}

One can thus define almost everywhere an action of
$G_n$ on $\mathbb R^{1, n-1}$ by means of the decomposition
(1.12), writing, for $t\in T_n$ and $g \in G_n$

\begin{equation}
tg = \gamma (t, g) t' \hskip.5cm; \hskip.5cm
\gamma(t, g) \in{\hat{\mathcal W}}_{n} , e' g=\varphi(t')
\end{equation}

Clearly, if $g = (\wedge, x) \in {\mathcal W}_n$ with $\wedge
\in Y_n, x \in T_n$, one has $t' = t\wedge + x$; if $g={\hat x}
= \hbox{exp}({\hat x}^\mu {\hat P}_\mu) \in {\hat T}_n$
one gets
\begin{equation}
t' = (t - (t^\mu {\hat x}_\mu) {\hat x})
( 1 - 2 t^\mu {\hat x}_\mu + q(t) q({\hat x}))^{-1}
\end{equation}

\noindent and $t'$ is defined when the denominator does not vanish. Elements
of ${\hat T}_n$ acting on $T_n$ are called
{\it{special conformal transformations}}.

\hskip1cm

 {\bf{c) Representations of ${\mathcal P}_n$ and $G_n$}}

\md

We are interested in determining which unitary irreducible
representations of ${\overline{\mathcal P}}_n$ can be extended to
${\overline{G}}_n$, or, conversely, which ones of ${\overline{G}}_n$
remain irreducible when restricted to ${\overline{\mathcal P}}_n$.
It will appear that they can all be realized as functional spaces
over the $n$-Minkowski space. We shall here begin by studying the UIR
of ${\overline{\mathcal P}_n}$ and operate a first selection among them;
in the next chapter we shall study the representations of ${\overline{G}_n}$
which satisfy the necessary constraints, and give a complete
description of the possible cases.

\md

Since ${\mathcal P}_n = {\mathcal L}_{n}\cdot T_n$ is a semidirect product
with abelian normal subgroup $T_n$, its UIR are determined by the
theory of Mackey \cite{Mack}: let $\mathcal O$ be an orbit of the dual of $T_n$
under the action of ${\mathcal L}_n$, and $\Gamma$ the stabilizer of a point in
$\mathcal O$; every UIR of ${\mathcal P}_n$ is equivalent to a representation
$U^S$ induced by a UIR  $S$ of $\Gamma$. Different orbits, or non-equivalent
representations of $\Gamma$ for the same orbit, induce non-equivalent UIR of
${\mathcal P}_n$.

\md

To construct $U^S$ one may proceed as follows: let $V$ be
the representation space of $S$; denote by
$x\longmapsto xh$ the action of ${\mathcal L}_n$ on $\mathcal O$,
with $x\in {\mathcal O}$ and $h\in {\mathcal L}_n$; let $\xi$ be the
point of $\mathcal O$ stabilized
by $\Gamma$, and let $x\longmapsto \tau_x$ be a smooth
injective mapping from $\mathcal O$ to ${\mathcal L}_n$, so that
 $x = \xi \tau_x$;
denote by $\gamma(x, h)$ the unique element of $\Gamma$ satisfying
$\gamma(x, h) \tau_{(x h)} = \tau_x h$; let $d\mu$ be a quasi-invariant
measure on $\mathcal O$ and let $\alpha$ be the positive function of
${\mathcal O}\times {\mathcal L}_n$ such that
 $d\mu(x h) = \alpha(x, h) d\mu(x)$. Let
${\mathcal H} = L^2 ({\mathcal O}, V, d \mu)$ be the Hilbert space of
$V$-valued functions $f$ such that:
$$ \int_{\mathcal O} \vert \vert f(x) \vert \vert^2_V\hskip.2cm
d\mu (x) < \infty.$$

For $h \in {\mathcal L}_n, t\in T_n$ and $f \in {\mathcal H}$
define $U^S$, acting on $\mathcal{H}$, by:
\begin{equation}
[U^S_{(h, t)} f] (x) = [\alpha(x,h)]^{1/2}. \hbox{exp}
[i g (x, t)] S_{\gamma(x, h)} f (xh)
\end{equation}
\noindent where $i^2 = -1$.

One thus has to determine the partition of the
$n$-Minkowski space into orbits under ${\mathcal L}_n$, and the corresponding
stabilizers and their UIR. For $n=4$ this was done
by E.P. Wigner, who also established all the theoretical
background neaded for this purpose, in a famous paper \cite{FlFr},
anterior to the formulation of the general theory by G. Mackey.
For other $n\geq 3$, the resolution into orbits is a straightforward
generalization of Wigner's results; it is given in Table 1.

\begin{center}
 {\bf{Table 1: Orbits and little groups for
the $n$-Poincar\'e group}}

\bigskip

\begin{tabular}{|c|c|c|c|}
\hline
Type & Orbit & Stabilized point & Stabilizer\\
\hline
0 & $x=0$ & 0 & ${\mathcal L}_n$\\
\hline
$\hbox{I}^\pm$ & $q (x) =0, \pm x_0 > 0$ & $\pm(e_0+e_{n -1})$ & $E_{n-2}$\\
\hline
$\hbox{II}^\pm_{|m|}$ & $q(x)=m^2>0, \pm x_0>0$ & $\pm|m| e_0$ & $Spin(n-1)$\\
\hline
$\hbox{III}_{|m|}$ & $q(x)=-m^2<0$ & $|m|e_{n-1}$ & $Spin(1, n-2)$\\
\hline
\end{tabular}
\end{center}

\hskip0.7cm

  In Table 1 the parameter $|m|$ runs over positive real numbers;
$Spin (1, n-2)$ denotes the twofold covering of $SO_0(1, n-2)$
(for $n=3$ this covering is merely $SO_0(1,1) \times \mathbb Z_2)$;
$E_{n-2}$ is the twofold covering of the
Euclidean group in $n-2$ dimensions, $Spin (n-2)\cdot
T_{n-2}$ (for $n=3$ this reduces to $\mathbb Z_2 \times T_1$).

\md

  One can immediately establish:

\noindent {\bf{Proposition 1.1:}} The UIR of
${\overline{\mathcal P}_n}$ corresponding
to orbits of types II, III and 0 (with the exception of the
trivial one) cannot be extended to UIR of ${\overline G_n}$.

\md
Proof: UIR of type 0 have trivial restriction
on $T_n$. Since ${\mathfrak g}_n = {\mathfrak t}_n
\oplus [{\hat{\mathfrak t}}_n, {\mathfrak t}_n]
\oplus [{\hat{\mathfrak t}}_n, [{\hat{\mathfrak t}}_n,
{\mathfrak t}_n]]$, the trivial representation of
${\overline {\mathcal P}_n}$ is the only possibility.

\md

Concerning types II and III, let $\widetilde{U}$ be the UIR
of ${\overline G_n}$ obtained by extending $U$. Since the parabolic subgroup
${\mathcal W}_n = Y_n. T_n$ contains ${\mathcal P}_n$, the restriction of
$\widetilde{U}$ on $ {\mathcal W}_n$ must also be irreducible. Since
${\mathcal W}_n$ is a semidirect product, its UIR are again
obtained by the orbit-stabilizer method. It turns out that
the orbits of $T_n$ under ${\mathcal W}_n$ are $x=0, x_\mu x^\mu = 0,
x_\mu x^\mu < 0$, and $x_\mu x^\mu > 0$, with
sign$(x_0)$ fixed if $x_\mu x^\mu \geq 0$;
thus, if $x_\mu x^\mu \neq 0$, the restriction of $\widetilde{U}$
to ${\mathcal P}_n$ is a direct integral, of representations over the
parameter $|m|$, which is not irreducible.

\hfill$\Box$

\md

So let us focus to UIR of type I, called
{\it{massless}} hereafter, by reference to the
$n=4$ case. Since ${\overline {\mathcal L}_n}$ is an invariant subgroup
of ${\overline Y_n}$, both groups acting on the same
homogeneous space ${\mathcal O}, E_{n-2}$ is an invariant
subgroup of $\Gamma '$, the stabilizer of $\xi$ in
${\overline Y_n}$, and ${\overline Y_n}/{\overline {\mathcal L}_n} = W
\times A \approx \mathbb Z_2 \times \mathbb R^{+\ast}$ is isomorphic
to $\Gamma '/E_{n-2}$.
More precisely one has

\begin{equation}
\Gamma ' = (W \times A' \times Spin
(n-2)) . T_{n-2}
\end{equation}

\noindent such that Lie $(T_{n-2})$ is generated by elements
$L_j = X_{j0} + X_{j, n-1}, 1 \leq j \leq n-2$;

\noindent Lie $(Spin (n-2))= \mathfrak{so}(n-2)$ is
generated by $X_{j, k}, 1 \leq j, k \leq n-2;
A' = \{\hbox{exp} t (X_{0,n-1} + X_{n, -1})\}_{t \in \mathbb R}
\approx \mathbb R^{+\ast}$ and $W=\{1, \hbox{exp} \pi (X_{0, -1} +
X_{n-1, n})\}$; $\Gamma '$ consists of the elements of
$G_n$ which commute whit
$P_0 + P_{n-1} = X_{0,-1} + X_{0,n} + X_{n-1, -1} +
X_{n-1, n}$.

\md

Let $S'$ be the inducing representation of $\Gamma '$
and $S$ its restriction to $E_{n-2}$, so that $U^S$ is the
restriction to ${\overline{\mathcal P}_n}$ of the representation $U^{S '}$
of ${\overline{\mathcal W}_n}$. Since $U^S$ must
be irreducible, $S$ must be irreducible too.

\md

To determine the UIR of $\Gamma '$, one can again
apply Mackey's theory of resolution into orbits. Without entering into many
details, one can see that $W \times Spin (n-2)$ stabilizes
the ``length'' $\ds x^2 = \sum^{n-2}_{j=1} (x_j)^2 =
-x_jx^j$ of an element $x$ of $T_{n-2}$, acting transitively on
the corresponding sphere.

\md

On the other hand, $\lambda \in A '$ acts as a dilatation
on $T_{n-2}$, sending $x$ to $\lambda x$. If $S '$
corresponds to a nonzero orbit, its restriction $S$ is a
direct integral of representations and $U^S$ is reducible. This leaves us
with:

\md

\noindent {\bf{Proposition 1.2:}} A necessary condition for a massless
representation $U^S$ of ${\overline{\mathcal P}_n}$ to extend to
${\overline G_n}$ is that the inducing representation $S$ is a
(finite-dimensional) UIR of \mbox{$Spin (n-2). T_{n-2}$} with trivial 
restriction to the normal subgroup $T_{n-2}$.
\hfill$\Box$

\md

For every such choice of $S$ and for either choice
of sign$(x_0)$, $U^S$ extends to $ {\overline{\mathcal W}_n}$, since
$S$ always extends to $S'$: one can always do this
by choosing a one-dimensional UIR of $A ' \times W$ the choice
being of course not unique. To see if the extension to
${\overline G_n}$ is possible, we shall use Lie algebraic methods.
Before proceeding further, we shall give the expression of the infinitesimal
operators of ${\overline{\mathcal P}_n}$, acting on a dense subspace of
analytic vectors of $\mathcal H$, the representation space of $U^S$.

\md

To be more precise about $\mathcal H$, the orbit $\mathcal O$ can be
parametrized by $\mathbb R^{n-1}-\{0\}:$ if $(x_0, \vec{x}) \in T_n$ is in
$\mathcal O$, let $\ds ||\vec{x}|| = (\sum_{\mu ' = 1}^{n-1} x_{\mu '}^2)
^{1/2}$. Since the orbit is massless, one has $x_0^2= ||\vec{x}||^2$, so that
if $\vec{x} \in \mathbb R^{n-1} - \{0\}$ is given, $x_0$ is fixed, its
sign being determined by the choice of $\mathcal O$. The quasi-invariant
measure $d\mu$ is defined by

\begin{equation}
d\mu (x) = d^{n-1} \vec{x} /||x||.
\end{equation}

 In fact $d\mu$ turns out to be invariant under the action
of ${\mathcal L}_n$ (but not under the action of dilatations), so that
the factor $\alpha$ in (1.17) equals 1. Putting
$S_{jk} = dS(X_{jk})$ acting on $V$, one obtains the
following expressions:

\begin{equation}
\left\{
\begin{array}{lclcl}
P_\mu & = &\sqrt{-1}\hskip.2cm x_\mu &   & \\
  &   &  &   &  \cr
X_{jk} & = & L_{jk} +  S_{jk} &,&1 \leq j, k \leq n-2\\
&   &  &   &  \cr
X_{j, n-1} &=& L_{j, n-1}+ B_j &,& 1\leq j\leq n-2\\
&   &  &   &  \cr
X_{0j} &=&x_0\partial_j + B_j&,&1\leq j\leq n-2\\
&   &  &   &  \cr
X_{0, n-1} &=& x_0 \partial_{n-1}&  & \\
\end{array}
\right.
\end{equation}

\noindent where

\begin{equation}
L_{\mu ' \nu'} = x_{\mu '} \partial_{\nu'} - x_{\nu'}
\partial_{\mu '},\qquad B_j = (x_0 + x_{n-1})^{-1}\sum^{n-2}_{k=1} x^k S_{jk}.
\end{equation}

We recall that we use the standard notation

\begin{equation}
\partial_{\mu '} = \partial / \partial x^{\mu '}
= - \partial/\partial x_{\mu '} \hskip1cm
(1 \leq \mu ' \leq n-1).
\end{equation}

This implies in particular:

\begin{equation}
[\partial_{\mu '}, x_0] = -x_{\mu '}/ x_0
\end{equation}

  It is clear that $U^S$ sends to zero the central element
$P^\mu P_\mu$ of ${\mathcal U}(\mathfrak{p}_n)$. This feature will be the
startpoint for the study of representations of ${\mathfrak g}_n$,
candidates to solve the problem.

\bigskip

\bigskip

\newpage

\section{Representations of $\mathfrak{so}(2,n)$ sending $P_\mu P^\mu$
to 0}\eqzero

\hskip1cm

 {\bf{a) Weight representations of $\mathfrak{so}(N)^{\mathbb C}$
and the Casimir element}}

\md

Let $g$ be a symmetric nondegenerate bilinear form
on $\mathbb R^N$, $I$ a set of cardinality $N, \{e_A\}_{A\in I}$
a basis of $\mathbb R^N$ and $g_{AB} = g(e_A,e_B)$. The orthogonal Lie
algebra $\mathfrak{g}= \mathfrak{so}(N, g)$ is spanned by generators
$X_{AB} = - X_{BA}$ such that

\begin{equation}
[X_{AB}, X_{CD}]= g_{BC}\hskip.2cm X_{AD} - g_{BD}\hskip.2cm X_{AC} -
g_{AC}\hskip.2cm X_{BD} + g_{AD}\hskip.2cm X_{BC}
\end{equation}

\noindent their action on $\mathbb R^N$ being (with bracket notations)

\begin{equation}
[X_{AB}, e_C] = e_A\hskip.2cm g_{BC} - e_B\hskip.2cm g_{AC}
\end{equation}

If $\{e^A\}$ is the dual basis, with $< e^A, e_B > = \delta^A_B$,
denoting by $g$ again the associated bilinear form on the dual,
with $g(e^A, e^B)= g^{AB}$, nondegeneracy implies $g^{AB}
g_{BC}= \delta^A_C$.

\md

We shall use the tensor $g$ for raising and lowering indices,
writing for instance ${X_A}^B$ for $X_{AC}\hskip.2cm g^{CB}$.

\md

The complexified Lie algebra $\mathfrak{g}^\mathbb C$
is independent of the choice of $g$ (up to isomorphism), the various real
forms being obtained by a suitable choice of the basis $\{e_A\}$,
fixing $\mathbb R^N$ in $\mathbb C^N$.

\md

We shall now introduce a Cartan subalgebra and a
Borel-type decomposition in $\mathfrak{g}^\mathbb C$
as follows:

\md

\noindent {\bf{Proposition 2.1:}} Let the indexing set $I$ be
$\{1, \ldots, N\}$
and assume $(g_{AA})^2 = (g^{AA})^2 = 1$, for every
$A\in I$. Fix the constant $\gamma$ by $\gamma =
 N/2-\mathrm{Rank}(\mathfrak{g})$, that is
$\gamma =0$ if $N$ is even and $\gamma ={\frac{1}{2}}$ if $N$ is odd. Let
$\hat{I} = \{\gamma + 1, \gamma + 2,\ldots, N/2\}$
be an indexing set of cardinality Rank ($\mathfrak{g}$);
let $q_A = g(e_A, e_A)$; for every
$a \in {\hat{I}}$, fix the constant $\eta_a$ such that

\begin{equation}
\eta^2_a = -q_{2a-1}\hskip.2cm q_{2a}
\end{equation}

\noindent(Lence $\eta_a^4=1$ and $\eta_a^\ast = \eta_a^{-1}
=\eta^3_a$) and define $H_a\in \mathfrak{g}^{\mathbb C}$ by:

\begin{equation}
H_a = \eta_a\hskip.2cm X_{2a-1,2a}
\end{equation}

The eigenvalues of $ad H_a$ are $0, +1, -1$;
for every index $A' \in I- \{2a -1, 2a\}$, the linear
combinations

\begin{equation}
\left\{
\begin{array}{lclcl}
X\left(\stackrel{+}{\hbox{\footnotesize\it{a}}}\right)_{A'}
& = & X_{2a, A'} & +& \eta_a q_{2a} X_{2a-1, A'}\\
&  &  &  &  \cr
X\left(\stackrel{-}{\hbox{\footnotesize\it{a}}}\right)_{A'}
& = & X_{2a-1, A'} & +&
\eta_a q_{2a-1} X_{2a, A'}\\
\end{array}
\right.
\end{equation}

\noindent are eigenvectors of $ad H_a$, satisfying:

\begin{equation}
\left\{
\begin{array}{clcl}
  &  [H_a, X\left(\stackrel{\pm}{\hbox{\footnotesize\it{a}}}
  \right)_{A'}] & =&
\pm\hskip.2cm X\left(\stackrel{\pm}{\hbox{\footnotesize\it{a}}}\right)_{A'}\\
  &\eta_a  [X\left(\stackrel{+}{\hbox{\footnotesize\it{a}}}\right)_{A'},
X\left(\stackrel{-}{\hbox{\footnotesize\it{a}}}\right)_{B'}] & =&
2\left(X_{A' B'} + g_{A' B'}
 H_a\right)\\
 & [X\left(\stackrel{+}{\hbox{\footnotesize\it{a}}}\right)_{A'},
X\left(\stackrel{+}{\hbox{\footnotesize\it{a}}}\right)_{B'}] & =&
 [X\left(\stackrel{-}{\hbox{\footnotesize\it{a}}}\right)_{A'},
X\left(\stackrel{-}{\hbox{\footnotesize\it{a}}}\right)_{B'}] = 0\cr
\end{array}
\right.
\end{equation}

Similarly the linear combinations $X\left(\stackrel
{\epsilon\hskip.2cm\epsilon'}
{\hbox{\footnotesize\it{a\hskip.2cm b}}}\right)$
defined by:

\begin{equation}
\left\{
\begin{array}{lclcl}
X\left({\stackrel{\varepsilon\hskip.2cm +}
{\hbox{\footnotesize\it{a\hskip.2cm b}}}}\right) & =&
X\left({\stackrel{\varepsilon}{\hbox{\footnotesize\it{a}}}}\right)_{2b} & +&
\eta_b\hskip.2cm q_{2b}\hskip.2cm X
\left({\stackrel{\varepsilon}{\hbox{\footnotesize\it{a}}}}\right)_{2b-1}\\
X\left({\stackrel{\varepsilon\hskip.2cm -}
{\hbox{\footnotesize\it{a\hskip.2cm b}}}}\right) & =&
X\left({\stackrel{\varepsilon}{\hbox{\footnotesize\it{a}}}}\right)_{2b-1} & +&
\eta_{b}\hskip.2cm q_{2b-1}\hskip.2cm X
\left({\stackrel{\varepsilon}
{\hbox{\footnotesize\it{a}}}}\right)_{2b}\\
\end{array}
\right.
\end{equation}

\noindent are simultaneous eigenvectors for every $ad H_c$,
belonging to the eigenvalue $\varepsilon 1$ if $c=a$, to
$\varepsilon '1$ if $c=b$ and to $0$ otherwise.

\md

Then:

1) The elements $H_a$ span a Cartan subalgebra $\mathfrak{h}$
of $\mathfrak{g}^\mathbb C$.

2) The set $\{X\left({\stackrel{\pm}{\hbox{\footnotesize\it{a}}}}\right)_{A'},
a \in {\hat{I}}, A' < 2a-1\}$ span a nilpotent subalgebra
$\mathfrak{n}^\pm$ of $\mathfrak{g}^\mathbb C$, for either
choice of the $\pm$ sign, such that
$\mathfrak{n}^- \oplus \mathfrak{h} \oplus
\mathfrak{n}^+$ is a Borel-type decomposition of
$\mathfrak{g}^\mathbb C$.

3) When $N$ is an even integer, all elements
$X\left(\stackrel{+ \hskip.2cm -}
{\hbox{\footnotesize\it{a\hskip.2cm b}}}\right)$ together with $\mathfrak{h}$
span a subalgebra $\mathfrak{l}$ isomorphic to $\mathfrak{gl} (N/2)$,
while elements $X\left(\stackrel{\pm \hskip.2cm \pm}
{\hbox{\footnotesize\it{a\hskip.2cm b}}}\right)$ span abelian
subalgebras $\mathfrak{n}^{\pm \pm}$, such
that $\mathfrak{l} \oplus (\mathfrak{n}^{++} \oplus
\mathfrak{n}^{--})$
is a Cartan decomposition of $\mathfrak{g}^\mathbb C$
corresponding to the real form $\mathfrak{so}^\ast (N)$.

4) A Cartan-Weyl basis of $\mathfrak{g}^\mathbb C$ is
$${\mathcal B}_0  = \left\{\frac{i}{2} \sqrt{\eta_a \eta_b}
\hskip.2cm X \left(\begin{array}{cc}
\varepsilon & \varepsilon'\cr
a & b \cr\end{array}\right) , \varepsilon = \pm , \varepsilon' =
\pm \right\}_{a<b}$$

\noindent if $N$ is even and

$${\mathcal B}_{1/2}={\mathcal B}_0 \cup  \left\{\sqrt{\frac{\eta_a}{q_1}}
X \left(\begin{array}{c}
\varepsilon\cr
a \end{array}\right)_1, \varepsilon = \pm \right\}_a$$

\noindent if $N$ is odd. Indeed one has, if $\{e_a\}_a$ is the dual basis of
$\{H_a\}_a$:

$$\begin{array}{l}
\left[\frac{i}{2} \sqrt{\eta_a \eta_b}\hskip.2cm X\left(
\begin{array}{cc}
+  & \pm\cr
a  & b \end{array}
\right), \frac{i}{2} \sqrt{\eta_a \eta_b}\hskip.2cm X\left(
\begin{array}{cc}
-  & \mp\cr
a  & b \end{array}
\right)\right] =  H_a \pm H_b\cr
 \cr
\left[ H, \frac{i}{2} \sqrt{\eta_a \eta_b}\hskip.2cm X\left(
\begin{array}{cc}
\pm  & \pm\cr
a  & b  \end{array}
\right)\right] = \pm (e_a + e_b) (H)
\frac{i}{2} \sqrt{\eta_a \eta_b}\hskip.2cm X\left(
\begin{array}{cc}
\pm & \pm\cr
  a & b \end{array}
\right)\cr
 \cr
\left[ H, \frac{i}{2} \sqrt{\eta_a \eta_b}\hskip.2cm X\left(
\begin{array}{cc}
\pm  & \mp\cr
 a & b \end{array}
\right)\right] = \pm (e_a - e_b) (H)
\frac{i}{2} \sqrt{\eta_a \eta_b}\hskip.2cm X\left(
\begin{array}{cc}
\pm & \mp\cr
  a & b \end{array}
\right)\end{array}$$

\noindent and, if $N$ is odd:
$$\begin{array}{l}
\left[\sqrt{\frac{\eta_a}{q_1}} X \left(\stackrel
{+}{\hbox{\footnotesize\it{a}}}\right)_1, \sqrt{\frac{\eta_a}{q_1}}
X\left(\stackrel
{-}{\hbox{\footnotesize\it{a}}}\right)_1 \right] = 2 H_a \cr
  \cr
\left[ H, \sqrt{\frac{\eta_a}{q_1}} X\left(\stackrel
{\pm}{\hbox{\footnotesize\it{a}}}\right)_1\right] = \pm e_a (H)
\sqrt{\frac{\eta_a}{q_1}} X\left(\stackrel
{\pm}{\hbox{\footnotesize\it{a}}}\right)_1.\end{array}$$

The root system is thus given by
$$\Delta_0 = \{\varepsilon e_a + \varepsilon' e_b, \varepsilon = \pm ,
\varepsilon' = \pm \}_{a<b}$$
if $N$ is even and
$$\Delta_{1/2} = \Delta_0 \cup \{\varepsilon e_a, \varepsilon = \pm \}_a$$
if $N$ is odd

\medskip

\noindent {\underline{Remark:}} If $q_{2a} = -q_{2a-1}$ then the
elements $H_a$ and $X\left(\displaystyle\begin{array}{c}
\pm \cr a  \end{array}\right)_{A'}$ belong to the real form $\mathfrak g$ from
which we started: with suitable modification of the indexing set $I$, one sees
that the dilatation operator or the Poincar\'e translations $P_\mu$,
imbedded in the conformal Lie algebra, are of this form. On the contrary,
such elements do not belong to the real form when $q_{2a} =  q_{2a-1}$ since
$\eta_a$ is imaginary.

\md

We are now interested in relating the eigenvalue of the Casimir
element $C$ of $\mathfrak{g}$, defined by

\begin{equation}
C = \frac{1}{2} X_{AB}\hskip.2cm g^{BC}\hskip.2cm X_{CD} g^{DA} = \frac{1}{2}
X_{AB} X^{BA}\end{equation}
to the extremal weight which defines a finite-dimensional irreducible
representation $\mathcal D$ of $\mathfrak{g}$.
Though this result is classical, we
shall give some details in view of further developments.

\md

So let $I = I' \cup I''$, with $I' = \{1, \ldots, N-2\}$
and $I''=\{N -1, N\}$. Splitting the summations one has:
\begin{equation}
C = C' + C'' + B
\end{equation}

\noindent where $C'$ is the Casimir element of $\mathfrak{so}(N-2)$ and $C''$
that of the complementary $\mathfrak{so}(2)$, that is
\begin{equation}
\begin{array}{lcl}
C'' &=& \frac{1}{2} (X_{N-1, N} X^{N, N-1} + X_{N, N-1} X^{N-1, N})\cr
    & &  \cr
    &=& -q^Nq^{N-1}(X_{N-1, N})^2\cr
    &  &  \cr
    &=& (\eta_{N/2} X_{N-1, N})^2 = (H_{N/2})^2
\end{array}\end{equation}

\noindent while

\begin{equation}B = -X_{A''A'} X^{A''A'}\end{equation}

\noindent where primed and double-primed indices are summed over
the sets $I'$ and $I''$ respectively.

\md

Develop now the expression $B^{-+}_{A' B'}$, symmetric
in the indices $A', B' \in I'$, defined as follows:

\begin{equation}\begin{array}{lcl}
B^{-+}_{A'B'} &=& \frac{1}{2} \eta_{N/2}
(X\left(\stackrel{-}{\hbox{\scriptsize\it{N/2}}}\right)_{A'}
X\left(\stackrel{+}{\hbox{\scriptsize\it{N/2}}}\right)_{B'} +
X\left(\stackrel{-}{\hbox{\scriptsize\it{N/2}}}\right)_{B'}
X\left(\stackrel{+}{\hbox{\scriptsize\it{N/2}}}\right)_{A'})\cr
&  & \cr
 &=&\frac{1}{2} \eta_{N/2}(X_{N-1, A'} + \eta_{N/2} q_{N-1} X_{N, A'})
(X_{N, B'} + \eta_{N/2} q_N X_{N-1, B'}) + \cr
&  & \cr
 & & \frac{1}{2} \eta_{N/2}(X_{N-1, B'} + \eta_{N/2} q_{N-1} X_{N, B'})
(X_{N, A'} + \eta_{N/2} q_N X_{N-1, A'})\cr
&  & \cr
 &=&\frac{1}{2}(\eta_{N/2}(X_{N-1, A'} X_{N, B'}- X_{N ,A'} X_{N-1, B'}) -\cr
 & & \cr
 & &(q^{N-1} X_{N-1, A'} X_{N-1, B'} + q^N X_{N, A'} X_{N, B'})) \cr
&  & \cr
 & & +\frac{1}{2}(\eta_{N/2}(X_{N-1, B'} X_{N, A'}- X_{, NB'} X_{N-1, A'}) -\cr
 & & \cr
 & &(q^{N-1} X_{N-1, B'}X_{N-1, A'} + q^N X_{N, B'}X_{N, A'}))\cr
 &  & \cr
 &=& - H_{N/2} g_{A' B'} - \frac{1}{2} g^{A''B''}(X_{A'' A'} X_{B'' B'} +
X_{A'' B'} X_{B'' A'})
\end{array}
\end{equation}

\md

Summing with $g^{A'B'}$ over the set $I'$ of cardinality
$N-2$ yields:

\begin{equation}
B = (N-2) H_{N/2} + B^{-+}
\end{equation}

\noindent where $B^{-+} = B^{-+}_{A'B'} g^{A'B'}$ (notice that the permutation
$N\longleftrightarrow N-1$ exchanges the $+$ and $-$ signs and
transforms $H_{N/2}$ to $-H_{N/2}$, while $B$ is left unchanged). Thus one has
\begin{equation}
C=H_{N/2}(H_{N/2}+N-2) + C' + B^{-+}
\end{equation}

Let now $V$ be a finite-dimensional irreducible
$\mathfrak{g}$-module, corresponding to the representation $\mathcal D$.
Let $s_{N/2}$ be the eigenvalue of ${\mathcal D}(H_{N/2})$ with maximal
real part, and let $V'$ be the subspace
\begin{equation}
V' = \{\varphi \in V; {\mathcal D} (H_{N/2}) \varphi = s_{N/2} \varphi\}
\end{equation}

If $N=2$, then $\dim V = \dim V' =1$ since the Lie algebra is abelian, and
${\mathcal D}(C) = s^2_1 = {\mathcal D}(H^2_1)$. If $N>2$, then
$V'\subseteq$ ker
${\mathcal D}(X\left(\stackrel{+}{\hbox{\scriptsize\it {N/2}}}\right)_{A'})$
 for every $A' \in I'$. It follows that $V'$ is an irreducible
${\mathfrak{g}}'$-submodule, where ${\mathfrak{g}}' = \mathfrak{so}(N-2)$ is
the subalgebra generated by $X_{A' B'}$ with $A', B' \in I'$ and also
${\mathcal D}(B^{-+}_{A' B'})$ vanishes on $V'$. Moreover, ${\mathcal D}$
 is integrable
to the compact real form since $V$ is finite dimensional, so that
${\mathcal D}(H_a)$ and $\pm {\mathcal D}(H_b)$ are conjugate for any choice of
$+$ or $-$ and of $a, b$ in $\hat{I}$: it can be proved that this implies
that every eigenvalue of ${\mathcal D}(H_a \pm H_b)$
(hence $2{\mathcal D}(H_a)$) is
an integer; in particular, $2 s_{N/2}\in \mathbb N$.

\md

If $N=3$, then ${\mathfrak{g}}' =\{0\}, C'=0, \dim V'=1$ and one gets the well
known formula

\begin{equation}
{\mathcal D}(C) = s(s+1), \hbox{ with } s=s_{3/2}
\end{equation}

If $N>3$, one may apply the same procedure to the ${\mathfrak{g}}'$-module
$V'$, introducing the maximal eigenvalue $s_{N/2 - 1}$ of
${\mathcal D}(H_{N/2 -1})$ restricted on $V'$, and so on.
Taking in account that $\vert s_a \vert\leq \vert s_{a+1}\vert$
because ${\mathcal D}(H_a)$ and ${\mathcal D}(H_b)$ are conjugate for every
$a$ and $b$, one easily gets by induction:

\noindent {\bf{Theorem 2.1:}} The extremal weight of an irreducible
finite-dimensional representation $\mathcal D$ of $\mathfrak{so}(N), N>2,$ is
determined by a sequence of positive numbers $s_a, a \in \hat{I}$, satisfying
 $s_{a+1} - s_a\in \mathbb N, 2s_a \in \mathbb N$, and such that

\begin{equation}
{\mathcal D}(C) = \sum^{N/2}_{a=\gamma+1} s_a(s_a + 2a -2).
\end{equation}

There is an extremal weight vector $\varphi \neq 0$, spanning
a one-dimensional subspace invariant by the Borel subalgebra
$\mathfrak{h} \oplus \mathfrak{n}$, such that
$${\mathcal D}(\mathfrak{n}^+) \varphi =\{0\}, {\mathcal D}(H_a)
\varphi = s_a \varphi \hbox{ if } a>1$$
and, when $N$ is an even integer, ${\mathcal D}(H_1)=\pm s_1 \varphi$
(representations with different choice of sign being inequivalent).
\hfill$\Box$

\md

One can also show that $\mathcal D$ is determined by the extremal weight up
to equivalence, and that the representation space is ${\mathcal D}({\mathcal U}
(\mathfrak{n}^-))\varphi$. We shall denote such a representation here after
by ${\mathcal D}(s_{N/2}, s_{N/2-1}, \ldots, s_{1+\gamma})$.

The corresponding Coxeter-Dynkin diagrams are:

\vspace{1cm}

%
%
%
%

\includegraphics{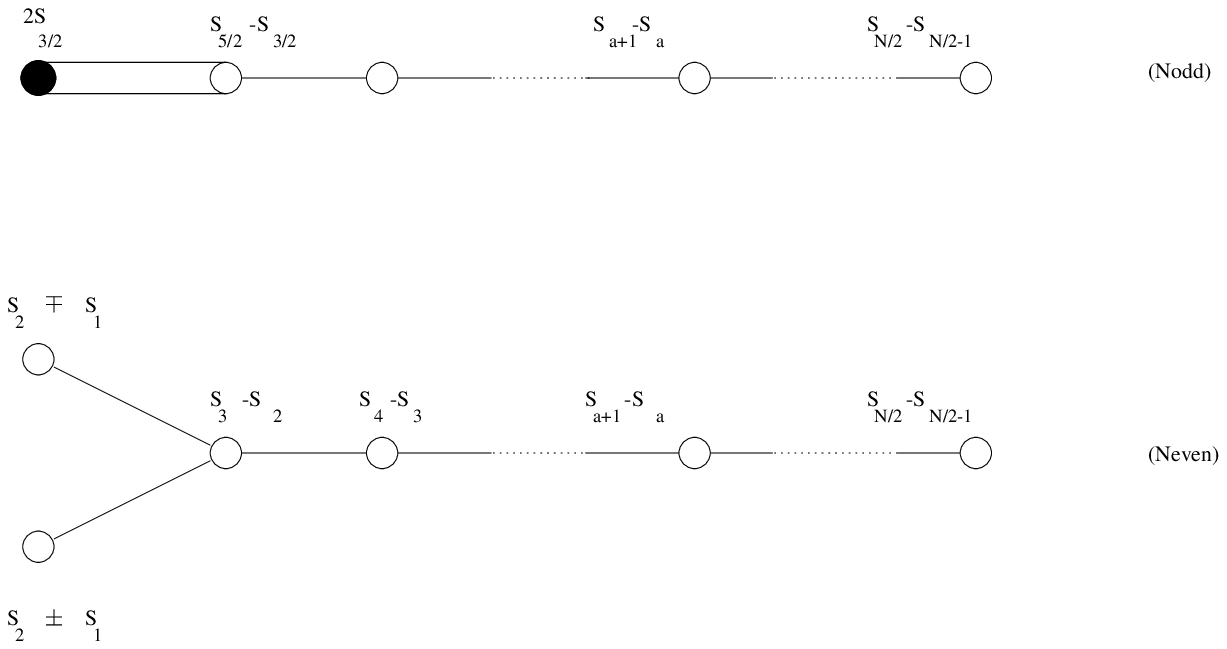}

\md

\noindent{\underline{Remark:}} Extremal weight representations of
$\mathfrak{g}$ with arbitrary range of the $s_a$'s can be defined, so that
(2.17) still holds: $\mathcal I$ being the left ideal
of ${\mathcal U}({\mathfrak{g}})$ corresponding to a one-dimensional
representation of $\mathfrak{h} \oplus \mathfrak{n}^+$, the left regular
representation on ${\mathcal U}({\mathfrak{g}})/{\mathcal I}$ has the desired
form. Integrability over some real form implies restrictions
on the range of $s_a$. In particular, for the real form
$\mathfrak{so}(2, N-2)$,
we shall denote by $d^{N-2, \varepsilon}_{(\alpha, \vec{s})}$ such a
representation, where $\varepsilon \in \{-1, +1\}$ and
$2 \alpha \not\in \mathbb N$;
the spectrum of
$d^{N-2, \varepsilon}_{(\alpha, \vec{s})}(-\varepsilon H_{N/2})$
is $\{\alpha - k, k \in \mathbb N\}$ and the eigenspace corresponding to the
 maximal eigenvalue $\alpha$ is an irreducible $\mathfrak{so}(N-2)$-module
corresponding to the weight $\vec{s} = (s_{(N-2)/2}, \ldots, s_{1+\gamma})$.

\bigskip

{\bf{b) Massless representations:}}

\medskip

Let us define the elements $\bar{F}_{AB}$ of the
enveloping algebra $\mathcal U$ of ${\mathfrak g}=
\mathfrak{so} (N)^\mathbb C$ by:

\begin{equation}\bar{F}_{AB} = \frac{1}{2}(X_{AC} g^{CD} X_{DB} +
 X_{BC} g^{CD} X_{DA}) = X_A \,^C X_{CB} - (\frac{1}{2} N-1) X_{AB}
\end{equation}
\noindent and the elements $F_{AB}$ as:

\begin{equation}
F_{AB} = \bar{F}_{AB} - \frac{1}{N} g_{AB} \bar{F}_{CD}
g^{CD} = \bar{F}_{AB} - \frac{2}{N} g^{AB} C
\end{equation}

The elements $F_{AB}$ are symmetric in the indices {\footnotesize\it{A, B}}
(as well as the $\bar{F}_{AB}$) and they span an irreducible
$\mathfrak{g}$-submodule $\mathcal F$ of $\mathfrak{g} \otimes \mathfrak{g}$
 under $ad\otimes ad$. For $N>2$ the dimension of ${\mathcal F}$ is
$N(N+1)/2 -1 = (N-1)(N+2)/2$ (for $N=2, {\mathcal F}$ is $\{0\}$);
$\mathcal F$ is isomorphic, as a $\mathfrak{g}$-module, to the Cartan subspace
$\mathfrak{p}$ in the Cartan decomposition
$\mathfrak{sl}(N) = \mathfrak{so}(N) \oplus \mathfrak{p}$ of
 $\mathfrak{sl}(N)$.

\md

Since $\mathcal F$ is irreducible, for every $Y \in {\mathcal F}$ the
two-sided ideal ${\mathcal U}\hskip.1cm Y\hskip.1cm  {\mathcal U}$ of
$\mathcal U$ contains $\mathcal F$; it follows:

\md

\noindent{\bf{Lemma 2.1: }}Given a representation $U$ of $\mathfrak g$,
if there is $Y \in{\mathcal F}$ such that $U(Y) = 0$, then
$U(Y') = 0$ for every $Y'$ in
${\mathcal U}\hskip.1cm Y\hskip.1cm {\mathcal U}$, and,
in particular, for every $Y' \in {\mathcal F}$.\hfill$\Box$

\md

Split now the indexing set $I$ into two disjoint sets $I'$ and
$I'' = \{S, T\}$. Let, as in the preceeding section, $\eta$ be such that
$\eta^2 = -g^{SS} g^{TT}$ and let $H= \eta X_{ST}$. The eigenvectors of $ad_H$
are given by:

\begin{equation}
X^+_{A'} = X_{SA'} - \eta q_S X_{TA'}; X^-_{A'} = X_{TA'} + \eta
q_T X_{SA'}
\end{equation}

\noindent for every $A'$ in $I'$.

\md

Summing over $A' \in I'$ these expressions one gets

\begin{equation}
\eta X^\pm_{A'} X^\mp_{B'} g^{A'B'} = -\eta^2 (q_T F_S + q_S F_T)
-2H^2 \pm (N-2) H+\frac{4}{N} C
\end{equation}

\noindent and

\begin{equation}
\begin{array}{lcl}
X^+_{A'} X^+_{B'} g^{A'B'} &=& F_{TT} - 2\eta q_T F_{ST} - \eta^2 F_{SS}\cr
   &   &  \cr
X^-_{A'} X^-_{B'} g^{A'B'} &=& F_{SS} - 2\eta q_S F_{ST} - \eta^2 F_{TT}
\end{array}
\end{equation}

One thus gets:

\noindent{\bf{Lemma 2.2: }} For every generator $X_{ST}$ with
$q^2_S = q_T^2 =1$, the expressions $X^\pm_{A'} X^\mp_{B'} g^{A'B'}$, in which
the summation runs over $I-\{S, T\}$ and the $X^\pm_{A'}$ are the eigenvectors
defined in (2.20), belong to $\mathcal F$.
In particular, if $N=n+2, I=\{-1, 0, 1, \ldots, n\}, \{ST\} = \{-1, n\}$,
the element $P_\mu P^\mu$ of the Poincar\'e enveloping algebra, canonically
imbedded in ${\mathcal U}(\mathfrak{so}(2,n))$, belongs to $\mathcal F$.
\hfill$\Box$

\md

{}From these two lemmas it follows:

\noindent{\bf{Proposition 2.2: }} If a representation $U$ of
${\mathcal U}(\mathfrak{so}(2,n))$ satisfies $U(P_\mu P^\mu) = 0$, then
$U$ vanishes on $\mathcal F$.
\hfill$\Box$

\md

Such a representation will be called {\bf{massless}} hereafter.

\md

We shall begin the study of massless representations by establishing:

\md

\noindent{\bf{Proposition 2.3: }} Let $U$ be a representation of $\mathfrak g$
which vanishes on $\mathcal F$. Let $N=N' + N''$ be any splitting of $N$ into
two positive integers, $I=I' \cup I''$ the corresponding
splitting of the indexing set, ${\mathfrak g}' = \mathfrak{so}(N')$ and
${\mathfrak g}'' = \mathfrak{so}(N'')$ the corresponding subalgebras.
Their Casimir elements $C'$
and $C''$ are related to the Casimir element $C$ of $\mathfrak g$ by:

\begin{equation}
U(C') - U(C'') = \frac{N' - N''}{N} U(C)
\end{equation}

In particular, if $N'' = 1$ and $I'' =\{1\}$ one has:

\begin{equation}
U(C') = \frac{N-2}{N} U(C)
\end{equation}

\begin{equation}
U(g^{AB} X_{1A} X_{B1}) = \frac{2q_1}{N} U(C)
\end{equation}

{\underline{Proof: }} Using distinct summations over $I'$, $I''$ and using
the definition of $F_{A'B'}$ one has

\noindent(2.26a)\centerline{$\displaystyle\begin{array}{rcl}
g^{A'B'} F_{A'B'} &=& g^{A'B'}(X_{A'C'}
g^{C'D'}X_{D'B'} + X_{A'A''} g^{A''B''} X
_{B'' B'} - \frac{2g_{A'B'}}{N}C)\cr
   &   &  \cr
 & =& 2C' + X_{A'A''} X_{B''B'} g^{A'B'} g^{A''B''}-
\frac{2N'}{N}C
\end{array}$}
\medskip

\noindent(2.26b)\centerline{$
\hskip-3.2cm g^{A''B''} F_{A''B''} =
 2C'' + X_{A''A'} X_{B'B''} g^{A'B'} g^{A''B''}- \frac{2N''}{N}C$}

\addtocounter{equation}{+1}
\medskip

\noindent and by substraction one gets the desired result,
since $U$ vanished on $\mathcal F$.\hfill$\Box$

\md

Let us now determine the irreducible massless representations.
Starting from low values of $N$, one first establishes:

\medskip

\noindent{\bf{Theorem 2.2:}} For $N=2$, every representation is massless,
$\mathcal F$ being $\{0\}$.

\noindent For $N=3$ the only irreducible massless representations are the
trivial and the spinorial (two-dimensional) one. For $N=4$, if
$\mathfrak{g} = \mathfrak{g}_1 \oplus \mathfrak{g}_2$ is the decomposition of
$\mathfrak{so}(4)$ into two ideals, each isomorphic to $\mathfrak{so}(3)$, an
irreducible representation is massless if and only if it vanishes on either
 $\mathfrak{g}_1$ or $\mathfrak{g}_2$.

\medskip

{\underline{Sketch of the proof: }} For $N=3$,
$\mathfrak{g} \otimes \mathfrak{g} =
{\mathcal F} \oplus \mathfrak{g} \oplus \mathbb C . C$, and one can show
(we leave this to the reader) that
$(C-\frac{3}{4}).\mathfrak{g}$ belongs to the ideal
${\mathcal U}\hskip.1cm{\mathcal F}\hskip.1cm{\mathcal U}$, so that
the quotient is a five-dimensional complex algebra, which turns out to be
$End_\mathbb C (\mathbb C^2) \oplus \mathbb C$.

\md

For $N=4$ one first sees that $\mathcal F$ is the span of all elements
$X_1 X_2$ with $X_i \in {\mathfrak g}_i$ sot that
${\mathcal U} {\mathcal F} = {\mathcal F} {\mathcal U}$ is
the intersection of the two maximal ideals $\mathfrak{g}_1 {\mathcal U}$ and
$\mathfrak{g}_2 {\mathcal U}$, hence the result.\hfill$\Box$

\md

So, from now on we shall suppose $N\geq 5$.

\md

Examining first the finite-dimensional case one gets:

\noindent {\bf{Theorem 2.3: }} A representation ${\mathcal D}(s_{N/2}, \ldots,
s_{1+ \gamma})$ is massless if and only if $\vert s_a \vert = s$ for every
$a\in \hat{I}$ where if $N$ is even (and $\gamma =0$) then $2s\in\mathbb N$
while if $N$ is odd $(\gamma = \frac{1}{2})$ then $s\in\{0, \frac{1}{2}\}$.
The corresponding value of the Casimir element is

\begin{equation}
C= \frac{1}{2} Ns(s+\frac{1}{2} N-1).
\end{equation}

Moreover, if $N$ is even, an extremal weight subspace carries a one dimensional
representation of the parabolic subgroup $\mathfrak{g}\mathfrak{l}(N/2) \oplus
\mathfrak{n}^{++}$, with trivial action of $\mathfrak{sl}(N/2)$ and
$\mathfrak{n}^{++}$.

\medskip

{\underline{Proof: }} We shall calculate $\bar{F}_{AB}$ on an extremal
vector $\varphi$. Using the notations of the preceeding section and taking
in account that $\mathfrak{n}^+$ vanished on $\varphi$, let $A, B< 2a-1$
for some  $a\in \hat{I}$; a calculation similar to (2.12) yields:
\begin{equation}
\sum_{i, j \in \{2a-1, 2a\}}\frac{1}{2} (X_{Ai} X_{jB} +
X_{Bi} X_{jA}) g^{ij} \varphi = H_a g_{AB}\varphi
\end{equation}

On the other hand, let $I'=\{1, \ldots, 2b\}$ and
$I''(b) =\{2b-1, 2b\} \subset I'$. Using distinct summations on primed and
double-primed indices, with $A', B' \in I'$ and $A'', B'' \in I''(b)$, one has,
using inductively (2.12):

\begin{equation}
g^{A''B''} X_{A''A'} X_{B'B''}G^{A'B'} \varphi = H_b(2H_b + 2b-2) \varphi
\end{equation}

\noindent hence

\begin{equation}
\sum_{A'', B''\in\{2b-1,2b\}}g^{A''B''}\bar{F}_{A''B''} \varphi =2[H_b(H_b+b-1)
+ \sum_{a>b} H_a]\varphi
\end{equation}

Since $F_{A'' B''}$ vanishes, one obtains:
\begin{equation}
\frac{2}{N}C\varphi= [H_b (H_b+ b-1) + \sum_{a>b} H_a]\varphi
\end{equation}

Equalling the expressions obtained for $b$ and $b+1$, one gets for consecutive
eigenvalues $s_b$ and $s_{b+1}$:

\begin{equation}
0 = s_b (s_b + b-1) -s_{b+1}(s_{b+1} +b-1) = (s_b -s_{b+1})
(s_b + s_{b+1} + b-1)
\end{equation}

For $b\geq1$ and $N$ odd or $b>1$ and $N$ even one has
$0 \leq s_b \leq s_{b+1}$ so that one must have $s_b = s_{b+1}$, and for $b=1$,
$N$ even, (2.32) becomes $\vert s_1\vert = \vert s_2\vert$, so that
$s=\vert s_a\vert$ is constant. For $b=N/2$, (2.31) gives the values of
the Casimir.

For $N$ odd one also has, by taking $A=B=1$ in (2.28) and
summing all over $a\in \hat{I}$:

\begin{equation}
\sum_{a\in\hat{I}} s_a = \frac{1}{2} (N-1)s=\frac{2}{N} C= s(s+\frac{1}{2} N-1)
\end{equation}hence $s(s-\frac{1}{2})=0$.

Notice also that $X\left(\stackrel{\varepsilon\hskip.2cm{\varepsilon'}}
{\hbox{\footnotesize\it{a\hskip.2cm b}}}\right) \varphi=0$
unless $\varepsilon = \varepsilon' = -$, because otherwise an eigenvalue equal
to $s+1$ would appear for some $H_a$, which is impossible. Since also $H_a-H_b$
vanishes on $\varphi, \varphi$ spans a one-dimensional representation
of $\mathfrak{g l}(N/2)\oplus \mathfrak{n}^{++}$ for even $N$, as stated.

\md

It remains to show every representation of this form is a
massless one. If $s=0$ we have the trivial one which is massless, and if
$s=\frac{1}{2}$ we have a spinorial representation $\mathcal D$ and Ker
$\mathcal D$ is a bilateral ideal of $\mathcal U$ containing $\mathcal F$;
this ends the odd $N$ case. For even $N$ and $s\geq 1$ one has
${\mathcal D}(\mathfrak{g}.\mathfrak{g}) \varphi =
({\mathcal D}(\mathfrak{n}^{--}.\mathfrak{n}^{--})
+ {\mathcal D}(\mathfrak{n}^{--})+\mathbb C)\varphi$.

\md

Diagonalizing the space $\mathcal F$ with respect to the Cartan subalgebra
$\mathfrak{h}$ one gets, among others, elements $F^{++}_a$ and $F^{--}_a$ such
that $[H_a, F_b^{\pm\pm}]=\pm 2 \delta_{ab} F^{\pm\pm}_b$, and all these
 elements are in Ker $\mathcal D$, since no elements of
$\mathfrak{n}^{--}.\mathfrak{n}^{--}$ or
$\mathfrak{n}^{--}$ have this property.
Writing $\mathfrak{h} =\mathfrak{h}' \oplus \mathbb C H$ with
$H=\sum_a H_a$ and $\mathfrak{h}' = \mathfrak{h} \cap \mathfrak{sl} (N/2)$,
one can substitute $\mathfrak{h}'$ with any conjugated subalgebra, and this
does not affect $\varphi$. The new elements $F^{\pm\pm}_b$ thus obtained are
distinct from the original ones, and as $\mathfrak{h}$ varies the whole of
$\mathcal F$ is spanned by such elements. It follows that
${\mathcal D}({\mathcal F}) \varphi = \{0\}$, and since
 ${\mathcal F} {\mathcal U} = {\mathcal U}{\mathcal F},
{\mathcal D}({\mathcal F})$ vanishes
on ${\mathcal D}({\mathcal U})\varphi$, so the representation
$\mathcal D$ is massless.\hfill$\Box$

\medskip

Consider now infinite-dimensional massless representations integrable to the
universal covering of the conformal group. Putting $n=N-2$, the maximal compact
subalgebra is $\mathfrak{k} = \mathfrak{so}(2) \oplus\mathfrak{so}(n)$, and the
 complexified Cartan subspace $\mathfrak{p}^\mathbb C$ is isomorphic to the
$\mathfrak{k}$-module $\mathbb C^2\otimes \mathbb C^n$. We shall again use the
usual notations for the $n$-conformal algebra, that is the indexing set will
be $I = I' \cup I''$ with $I' =\{1, \ldots, n\}, I'' = \{-1, 0\}$ and the
indexing set $\hat{I}$, for the Cartan subalgebra,
$\{0, \frac{n}{2}, \frac{n}{2}-1, \ldots\}$, we shall denote by $H_0$ the
 central element $\eta X_{-1, 0}$ (with $\eta^2 = -1$ and
$g_{-1, -1} = g_{00} =1$) of $\mathfrak{k}^\mathbb C$.

\medskip

The space $\mathcal H$ of the representation $U$ is a direct sum of
 $\mathfrak{k}$
submodules $W(s_0, \vec{s})$, where $s_0$ is the eigenvalue of $H_0$ and
$\vec{s}$ the extremal weight of $\mathfrak{so}(n).$ $\mathfrak{p}^\mathbb C$
acts on $W(s_0,\vec{s})$ like
$(\mathbb C^2\otimes \mathbb C^{n}) \otimes W(s_0, \vec{s})$:
this tensor product splits in general into $2n$ components
$W(s_0 + \varepsilon, \vec{s} + \Delta \vec{s})$ with $\varepsilon = \pm 1$
and $\Delta s_a = (\Delta \vec{s})_a = \pm 1$ for one $a \in \hat{I} - \{0\}$
(at most if $n$ is odd, exactly if $n$ is even), all remaining coordinates
of $\Delta \vec{s}$ being $0$ (if $n$ is odd $\Delta \vec{s} = \vec{0}$ also
exists in general). When $\Delta s_a = \pm 1$ and $s_{a+1}=s_a$ the
corresponding component vanishes, since the resulting weight would not respect
the ordering $s_{a+1} \geq s_a$.
In particular, $\mathbb C^n \otimes W$ always contains a component
$W^\uparrow$ for  which $\Delta s_{n/2}=1$ (the maximal eigenvalue increases)
and a component $W^\downarrow$ for which $\Delta s_{n/2} = -1$; this latter is
 nonzero only if $s_{n/2}-1 \geq s_{n/2-1}$.

\medskip

Assume now $U$ irreducible and massless and take
 $\varphi$ in $W(s_0, \vec{s})$.
Because of (2.23) $s_0$ is related to the Casimir $C'$ of $\mathfrak{so}(n)$ by

\begin{equation}
(C'-s_0^2) \varphi = \frac{n-2}{n+2} C\varphi
\end{equation}

Let $\vert s_0\vert = \varepsilon s_0$. For $\varphi$ in $W(s_0, \vec{s})$
one has:

\begin{equation}
[H^2_0, X\left(\stackrel{\pm \varepsilon}{0}\right)_{A'}] \varphi = (\pm 2
\varepsilon s_0 +1) X\left(\stackrel{\pm \varepsilon}{0}\right)_{A'} \varphi
\end{equation}

On the other hand, if $\varphi$ is an extremal vector then
$X\left(\stackrel{+}{\hbox{\footnotesize\it{n/2}}}\right)_{A''} \varphi$
\hskip.2cm  belongs to $W(s_0, \vec{s})^\uparrow$
(with $A'' \in \{-1, 0\} = I'')$ since the maximal eigenvalue
increases, so that, by (2.17):
\begin{equation}
[C', X\left(\stackrel{+}{\hbox{\footnotesize\it{n/2}}}\right)_{A''}] \varphi =
(2 s_{n/2} + n-1) X\left(\stackrel{+}{\hbox{\footnotesize\it{n/2}}}
\right)_{A''}\varphi
\end{equation}

Since the difference $C'-H_0^2$ is constant, these two equations imply:

\begin{equation}
(s_{n/2} + n/2 -1 \mp \vert s_0\vert) X
\left({\stackrel{\pm \varepsilon}{0}} \hskip.2cm{\stackrel{+}
{\hbox{\footnotesize\it{n/2}}}}\right)
\varphi =0
\end{equation}

\noindent hence $X\left(\stackrel{-\varepsilon}{0}\hskip.2cm
\stackrel{+}{\hbox{\footnotesize\it{n/2}}}\right)$ vanishes on $\varphi;
X\left(\stackrel{+\varepsilon}{0}\hskip.2cm\stackrel{+}
{\hbox{\footnotesize\it{n/2}}}\right)\varphi$
is an extremal vector of
$W(s_0, \vec{s})^\uparrow$ and the only non-vanishing component of
$\displaystyle\sum_{A' \in I-\{n-1, n\}}\lambda^{A'} X
\left(\stackrel{+}{\hbox{\footnotesize\it{n/2}}}\right)_{A'}\varphi$,
so it is nonzero (otherwise $s_{n/2}$, hence $C'$, would be
bounded and $U$ would be finite dimensional),and we get:

\begin{equation}
\vert s_0\vert = s_{n/2} + n/2 - 1
\end{equation}

\noindent for every $W(s_0, \vec{s})$. It also follows that

\begin{equation}
X\left(\stackrel{-\varepsilon}{0}\right)_{A'} W(s_0, \vec{s})
\subset W(s_0,\vec{s})^\downarrow
\end{equation}

\noindent and one can transform (2.21) to:

\medskip

\noindent(2.40a)\centerline{\hskip-1.5cm$X\left(\stackrel{-}
{\hbox{\footnotesize\it{n/2}}}
\hskip.2cm\stackrel{-\varepsilon}{0}\right)
X\left(\stackrel{+}{\hbox{\footnotesize\it{n/2}}}\hskip.2cm
\stackrel{\varepsilon}{0}\right)
\varphi = 4 (\frac{2}{n+2} C-s_{n/2}(s_{n/2} + n/2))\varphi$}

\medskip

\noindent(2.40b)
\centerline{$X\left(\stackrel{+}
{\hbox{\footnotesize\it{n/2}}}\hskip.2cm \stackrel{\varepsilon}{0}\right)
X\left(\stackrel{-}{\hbox{\footnotesize\it{n/2}}}\hskip.2cm
\stackrel{-\varepsilon}{0}\right)
\varphi = 4 (\frac{2}{n+2} C-(s_{n/2}-1) (s_{n/2}-1 + n/2))\varphi$}

\addtocounter{equation}{+1}
\medskip

One also checks that $X\left(\stackrel{-}{\hbox{\footnotesize\it{n/2}}}
\hskip.2cm \stackrel{-\varepsilon}{0}\right)\varphi$ is the only
nonvanishing component in

\noindent$\displaystyle\sum_{A'\in I'} \lambda^{A'}
X\left(\stackrel{-\varepsilon}{0}\right)_{A'} \varphi$ (otherwise  an
eigenvalue of $H_{n/2}$ superior to $s_{n/2}-1$ would appear), and it is again
an extremal vector. When $s_{n/2}$ reaches its minimal value, $s$, every
$X\left(\stackrel{-\varepsilon}{0}\right)_{A'} \varphi$ is zero and (2.40b)
gives

\begin{equation}
2C=(n+2)(s-1)(s-1 + n/2) =(n+2). \hbox{ Inf }\vert s_0\vert.(\hbox{ Inf}
\vert s_0\vert-\frac{n}{2}).
\end{equation}

It follows that $-\varepsilon H_0$ has a negative maximal value equal to
$-(s-1 + n/2)$; an extremal vector $\varphi$ for $s_{n/2} = s$ is an extremal
vector for the whole representation space and the nilpotent subalgebra
$\mathfrak{n}^+$ vanishes on $\varphi$. As for the remaining coordinates of
$\vec{s}$, one easily sees that they are all equal to $s$ (or $-s$ for the
last one for even $n$), and that $s=0$ or $1/2$ when $n$ is odd, the proof
being exactly the same as in the finite-dimensional case.

\md

Using the notations of Theorem 2.1 and the Remark following it, one can
 summarize:

\md

\noindent{\bf{Theorem 2.4: }} Every infinite-dimensional
irreducible massless representation of $\mathfrak{so}(2,n)$,
for $n\geq 3$, integrable to $\overline{G}_n$, is a weight representation
$d_{(-(s+n/2 -1), \vec{s})}^{n, \varepsilon}$, ${\mathcal D}(\vec{s})$ being
itself a massless representation of $Spin(n)$, that is $\vert s_a\vert = s$
for every $a$. The eigenspace of $\varepsilon H_0$ corresponding to the
eigenvalue $(s + n/2 -1 + k), k\in \mathbb N$, is an irreducible
$\mathfrak{so}(n)$-module corresponding to the representation
${\mathcal D}(\vec{s} + (k, 0, \ldots, 0))$. The values of the Casimir element
$C$ is given by (2.41).\hfill$\Box$

\medskip

In addition one has:

\noindent{\bf{Proposition 2.4: }} The massless representations
$d_{(-(s+\frac{1}{2} n-1), \vec{s})}^{n,\varepsilon}$ are integrable to unitary
representations of $\bar{G}_n$.

\medskip

{\underline{Proof:}} From what precedes, every $\mathfrak{so}(n)$-submodule
$W_k$ has multiplicity one, it carries the representation
 ${\mathcal D}(\vec{s} +
(k, 0, \ldots,0))$ of $\mathfrak{so}(n)$, and the unique eigenvalue of
$\varepsilon H_0$ on it is  $s + k +\frac{1}{2}n-1 (k\in\mathbb N)$.
Since there is a natural $\mathfrak{k}$-invariant scalar product on
each $W_k$ and since $\mathfrak{p}. W_k \subset W_{k-1}\oplus W_{k+1} $, it
is sufficient to show that
$\vert \vert X\varphi \vert \vert^2 = q(X) \vert \vert \varphi \vert \vert^2$
for every $X \in \mathfrak{p}$ such that $[\varepsilon H_0, X] = \pm X$,  with
$q(X) \geq 0$; it is clear that $q(X)$ belongs to the spectrum of $X^\ast X$.

\medskip

There is no loss of generality in assuming that $\varphi$ is an extremal vector
 of $W_k$; but then $X$ must be proportional to either $X_+ =
X\left(\stackrel{+}{\hbox{\footnotesize\it{n/2}}}
\hskip.2cm\stackrel{\varepsilon}{0}\right)$ or
$X_-=X\left(\stackrel{-}{\hbox{\footnotesize\it{n/2}}}
\hskip.2cm\stackrel{-\varepsilon}{0}\right)$,
with $(X_\pm)^\ast = -X_\mp$
and where $X_\pm \varphi$ is an extremal vector of $W_{k\pm 1}$; from (2.40)
one sees that $X^\ast X$ is scalar, with
$$\begin{array}{lcl}
X^\ast_+ X_+ &=& (s+k)(s+k+\frac{n}{2}) -
(s-1)(s-1 +\frac{n}{2})=(k+1)(2s + k-1+\frac{n}{2})\cr
 &  &  \cr
X^\ast_- X_- &=& (s+k-1)(s+k-1+\frac{n}{2}) -
(s-1)(s-1 +\frac{n}{2})= k (2s + k-2+\frac{n}{2})
\end{array}$$

\noindent and these expressions are positive
for $k\in \mathbb N, s\geq 0$ and $n \geq 3$.\hfill$\Box$

\bigskip


\noindent{\bf{c) Conformal imbedding of Poincar\'e massless representations}}

\md

Having determined all possible candidates, up to equivalence, we shall now
examine whether a massless representation $U$ of Poincar\'e extends to one of
them, and how.

We shall proceed by combining the expressions of the generators of
${\mathcal P}_n$
given in (1.20) to obtain elements of the ideal ${\mathcal U}{\mathcal F}$.
 One first establishes:

\medskip

\noindent{\bf{Proposition 2.5: }}Given the expressions (1.20) of the Poincar\'e
 generators
of $U$, if $P_\mu$ is identified with $X_{\mu, -1} + X_{\mu,n}$ (with $\mu \in
J=\{0, \ldots, n-1\}$), then the dilatation operator $D =X_{n,-1}$, satisfying
 $[D,P_\mu] = P_\mu$, is given by

\begin{equation}
D= x_{\mu '} \partial^{\mu '} + (n-2)/2,\hskip.3cm \mu' \in \{1, \ldots, n-1\}.
\end{equation}

{\underline{Proof:}} Using a summation index $\lambda \in J$, and since
$g_{-1,-1} = -g_{n, n} = 1$, one has:

\begin{equation}\begin{array}{lcl}
F_{-1\mu} + F_{n \mu} &=& (X_{-1\lambda} + X_{n\lambda}) {X_\mu} ^\lambda +
(X_{n, -1} X_{-1, \mu} -X_{-1, n} X_{n, \mu}) -\frac{1}{2} n
(X_{-1, \mu} + X_{n \mu})\cr
  &  &  \cr
 &=& P_\lambda({X_\mu}^\lambda + (D+1 - \frac{n}{2}) \delta^\lambda_\mu)
\end{array}
\end{equation}

\noindent substituting the expressions of the generators,
and putting $F_{AB}=0$, one gets, for every $\mu$ in $J$:

\begin{equation}
0 = x_\mu (D+1-\frac{n}{2} -x^{\mu'}\partial_{\mu'})
\end{equation}

\noindent hence the result announced.\hfill$\Box$

\medskip

Now, one can rewrite (2.23) as:

\begin{equation}
\frac{1}{2} X_{\lambda \mu} X^{\mu \lambda} = \frac{n-2}{n+2} C + D^2
\hskip.3cm (\hbox{ mod }{\mathcal U}{\mathcal F})
\end{equation}

\noindent and one also has

\begin{equation}
\frac{1}{2}(P_\mu \hat{P}_\nu + P_\nu \hat{P}_\mu) =
 {X_\mu}^\lambda X_{\lambda \nu}-\frac{1}{2}
(n-2) X_{\mu \nu}-g_{\mu \nu}(D+\frac{2}{n+2} C) -F_{\mu \nu}
\end{equation}

\noindent where $\hat{P}_\nu = X_{\nu, -1} - X_{\nu n}$,
satisfying $[D, \hat{P}_\nu] = -\hat{P}_\nu$.

\md

Substituting the expressions of the generators in (2.45)
and (2.46), one obtains,
after some calculations which we do not reproduce:

\begin{equation}
C = \frac{n+2}{n-2} C'' - \frac{1}{4}(n+2)(n-2)
\end{equation}

\noindent where $C''=\frac{1}{2} S_{ij} S^{ij}(i,j \in\{1,\ldots,
n-2\})$ is the Casimir element of the inducing representation
$S$; from (2.46) one gets expressions of the form

\begin{equation}
P_\mu \hat{P}_\nu + P_\nu \hat{P}_\mu = x_\nu G_\mu + x_\mu G_\nu +
E_{\mu \nu}
\end{equation}

\noindent with

\medskip

\noindent(2.49a)\hskip-1.2cm\centerline{$E_{ik} = (S_{ij} S^j_k +
S_{kj} S^j_i) -\frac{4}{n-2} g_{ik} {C''}; i, j, k\in\{1, \ldots, n-2\}$}

\vskip.5cm

\noindent(2.49b)\centerline{$E_{\alpha k} = \sigma
(\alpha)x^i E_{ik}(x_0 +x_{n-1})^{-1}
; \alpha \in\{0, n-1\}, \sigma (0) = -\sigma(n-1) = -1$}

\vskip.5cm

\noindent(2.49c)\hskip-1.4cm\centerline{$E_{\alpha \beta} =
 \sigma (\alpha) \sigma(\beta) x^i x^k E_{ik}(x_0 +x_{n-1})^{-2};
\alpha, \beta \in\{0, n-1\}$}

\addtocounter{equation}{+1}

\medskip

Since $P_\mu = \sqrt{-1}\hskip.2cm x_\mu$, the consistency of the
$\frac{n(n+1)}{2}$ equations (2.48) implies that $E_{\mu \nu} =0$;
in particular $E_{ik}=0$, that is $S$ is a massless representation of the
little group \mbox{$Spin (n-2)$}. Carrying out the calculations, one finally
 obtains:

\md

\noindent{\bf{Theorem 2.5: }} A massless representation of
$\bar{\mathcal P}_n (n \geq 3)$ induced by the representation $S$ of
 $Spin(n-2).T_{n-2}$ (trivial on $T_{n-2}$)
extends to a massless UIR of $\bar{G}_n$ iff
$S$ itself is massless, that is of the form
 ${\mathcal D} (s, \ldots, s,\pm s), 2s
\in \mathbb N$, if $n$ is even and of the form
${\mathcal D}(s, \ldots, s), s=0$ or
$ \frac{1}{2}$, if $n$ is odd. The extension is unique, the form of the
remaining generators (of $\mathfrak{g}_n$) being completely determined by
those of $\mathfrak{p}_n$ in (1.20): $X_{n,-1} = D$ is given by (2.42) and
$\hat{P}_\mu$ by:

\begin{equation}
\sqrt{-1} \hat{P}_\mu = x_\mu  \Delta + 2 (x_0 + x_{n-1})^{-1} D_\mu +
2 D \partial_\mu
\end{equation}

\noindent with $\displaystyle \partial_0 = 0, \Delta = \sum^{n-1}_{j=1}
\partial^2_j$ and:

\noindent(2.50a)\centerline{$D_j = (L_{0k} - L_{k, n-1}) S^k j\hskip.4cm
(j, k \in\{1, \ldots, n-2\})$}

\vskip.5cm

\noindent(2.50b)\centerline{$D_{n-1} = -D_0 =\frac{1}{2}L^{jk} S_{kj} +
s(s+\frac{1}{2} n-2)$}

\vskip.5cm

\noindent and the values of the Casimir element for the inducing and the
extended representations are:

\begin{equation}
C'' = \frac{1}{2}(n-2) s(s+\frac{1}{2} n-2);\hskip.3cm C=
\frac{1}{2}(n+2)(s-1)(s+
\frac{1}{2} n-1)
\end{equation}
\hfill$\Box$

{\underline{Remark: }} The constraints upon $S$ are relevant for $n>4$. Indeed,
for the classical case, $n=4$, the little group is $SO(2). T_2$, and the
elements $E_{ij}$ in (2.49) are identically zero: every such representation
extends to the conformal group, as shown in \cite{AFFS}.
For $n=3, \mathfrak{so}(n-2)=\{0\}$ and all elements $S_{ij}$ vanish; notice
that $C''$ vanishes in (2.51) for $n=3$ and for either $s=0$ or
$s=\frac{1}{2}$. However, the choice of $S$ is relevant: it corresponds to the
inducing representation of $Spin(1) = \{1, -1\}$ and determines whether the
center of $Spin (3)= SU(2)$ is trivially represented $(s=0)$ or not
$(s =\frac{1}{2})$, the lowest $\mathfrak{so}(3)$-module occuring in the
representation space having dimension $2s+1$.

Now, for given $s$, there are two possible choices for the extension,
$d^{n,\varepsilon}_{(-(s+\frac{1}{2} n-1), \vec{s})}$, such that the spectrum
of $\varepsilon \sqrt{-1}\hskip.2cm X_{-1, 0}$ is positive, so that it remains
to identify which one is obtained. We shall show:

\md

\noindent {\bf{Proposition 2.6: }} For a given sign $\varepsilon$ of
$x_0=\varepsilon. \vert x_0\vert$, the representation $U^S$
of $\bar{\mathcal P}_n$ extends to
 $d^{n, \varepsilon}_{(-(s + \frac{1}{2} n-1), \vec{s})}$.

\md

{\underline{Proof: }} On every $\mathfrak{so}(3)$-submodule $W_k$ the absolute
value of $s_0$ is $s+k+\frac{1}{2} n-1$, while the eigenvalues of $H_{n/2}$
run from $-(s+k)$ to $s+k$, so that the spectrum of $E=2\sqrt{-1}(X_{-1, 0} +
 X_{n-1, n})$ has the same sign as $H_0$ and a lowest element equal,
in absolute value, to $n-2$.
Substituting with the differential operators obtained one gets:

\begin{equation*}
\begin{array}{lcl}
E &=& \sqrt{-1} (-\hat{P}_0 - \hat{P}_{n-1} -
P_0 + P_{n-1})\cr
  &  &  \cr
 &=& -(x_0 + x_{n-1}) \Delta - 2 D \partial_{n-1} + (x_0 - x_{n-1}),
\hskip.3cm D= x_{\mu'} \partial^{u'} + \frac{n-2}{2}
\end{array}
\end{equation*}

Take $f \in {\mathcal H}$ so that $f$ depends only on
$x_0 = \varepsilon (- x^{\mu'} x_{\mu'})^{1/2}$, and denote by $d_0$ the
differential operator $\frac{d}{dx_0}$. For such an $f$ one has:

$$Ef = ((x_0 - x_{n-1})(1-d^2_0) -(n-2) d_0)f.$$

If ${d_0}^2 f = f$, that is, for example, if
$f(x_0)=e^{\pm \varepsilon x_0} v$, with $v\in V$, one gets:
$$Ef = \mp \varepsilon (n-2)f$$

Since only $e^{-\varepsilon x_0} v$ is a square-integrable function from
$\mathbb R^{n-1}$ to $V, \varepsilon E$ has a positive spectrum, and so does
$\varepsilon H_0$, hence the desired result.\hfill$\Box$

\md

{\underline{Remark: }} When $S$ is trivial, the Fourrier transform on
$\mathcal H$
sends it on the subspace $\hat{\mathcal H}$ of $L^2(\mathbb R^{1,n-1}, d\mu)$
which is the closure of all analytic functions satisfying $\partial_\mu
\partial^\mu f =0$. The action of $\bar{G}_n/\bar{\mathcal P}_n$ on
$\hat{\mathcal H}$
is obtained from the action of dilatations and special conformal
transformations on the $n$-Minkowski space.
What we have shown is that, for $S$ acting on $V$, the representation
$U^S$ acting on $\hat{\mathcal H} \otimes V$ can be extended iff $S$ is
massless.
\newpage


\section{Massless representations and the De Sitter groups}\eqzero

\noindent{\bf{a) Subgroups of $G_n$}}

Let $x \in \mathbb R^{n+2}$. If its quadratic form $q(x)$ is
positive (negative), its stabilizer ${\mathcal S}_n(x)$ is isomorphic to
 $SO_0(1,n)) (SO_0(2, n-1))$. For distinct choices of $x$,
${\mathcal S}_n(x)$ and
${\mathcal S}_n(x')$ are conjugated subgroups iff $q(x). q(x')>0$, so we shall
denote them by ${\mathcal S}^\pm_n$ (for $q(x) = \pm \vert q(x)\vert$) and
call them the {\it{$n$-De Sitter subgroups}} of real rank 1  or 2 respectively,
in analogy with the classical case $n=4$. Clearly, ${\mathcal S}^-_n =G_{n-1}$.
When $q(x)=0$ the stabilizer is isomorphic to ${\mathcal P}_n$.

\md

We shall examine here the restriction to the twofold covering $\bar{\mathcal
S}^\pm_n$ of a massless representation $d$ of $\bar{G}_n$, establishing that it
is either irreducible, or the direct sum of two factors. Also, since ${\mathcal
P}_n$ is a Wigner-Inon\"{u} contraction of ${\mathcal S}^\pm_n$, we shall
establish that the restriction of $d$ on $\bar{\mathcal S}^\pm_n$ can be
contracted to its
restriction on $\bar{\mathcal P}_n$.

\bigskip

\noindent{\bf{b) Restriction  of $d^{n, \varepsilon}_{(-(s+\frac{n}{2}-1),
\vec{s})}$ to ${\mathcal S}^\pm_n$}}

We have already established that the Casimir element $C'$ of
$\mathcal{S}^\pm_n$ is scalar and equal to $C.(N-1)/N,$ in (2.27).
We shall next continue with

{\bf{Lemma 3.1: }} Let $\mathfrak{g}'$ be the Lie algebra of
${\mathcal S}^\pm_n$, ${\mathcal U}'$ its enveloping algebra, and let
$e_{w}, w\in I$, be a basis vector
stabilized by $\mathfrak{g}'$. Let $d$ be a massless representation of
$\mathfrak{g}$ acting on a Hilbert space $\mathcal{H}$ and
$W'$ be a $\mathfrak{g}'\cap \mathfrak{k}$ invariant subspace of a
$\mathfrak{k}$-type $W$. Let $V_0, V_1$ be the prehilbert spaces

\begin{equation}
V_0 = d({\mathcal U}') W';\hskip.3cm V_1 = \sum_A d({\mathcal U}' X_{Aw})W'
\end{equation}

\noindent and ${\mathcal H}_0, {\mathcal H}_1$ their closures. Then either
${\mathcal H} = {\mathcal H}_0 = {\mathcal H}_1$ or
${\mathcal H} = {\mathcal H}_0 \oplus {\mathcal H}_1$.

\medskip

{\underline{Proof: }} Let $\mathfrak{x}$ the $\mathfrak{g}'$-invariant
subspace of $\mathfrak{g}$ spanned by the generators $X_{Aw}$, such that
$\mathfrak{g} = \mathfrak{x}\oplus \mathfrak{g}'$.
Since ${\mathcal U}' = \oplus_{k \in \mathbb N} \hskip.2cm {\mathcal U}' S^{k}
(\mathfrak{x})$, where $S^k(\mathfrak{x})$ contains the fully symmetrized
polynomials of degree $k$ in the generators of $\mathfrak{x}$, it is
sufficient to show that $d$ sends $S^2(\mathfrak{x})$ to
$S^0(\mathfrak{x})  =  \mathbb C$. But one has:

\begin{equation}
X_{A'w} X_{B'w} + X_{B'w} X_{A'w} = g_{ww} ({X_{A'}}^{D'} X_{D'B} +
{X_{B'}}^{D'} X_{D'A'})-2\bar{F}_{A'B'})
\end{equation}

\noindent and since $d(\bar{F}_{A'B'})= g_{A'B'} 2C/N \in
\mathbb C, d$ sends $S^2(\mathfrak{x})$
to ${\mathcal U}'$.\hfill$\Box$

\md

Now, if $G'={\mathcal S}^+_n, \mathfrak{x}=\{\lambda^A X_{-1, A}\},
\mathfrak{g}'\cap \mathfrak{k} = \mathfrak{so}(n)$, and the
$\mathfrak{k}$-type $W(k)$ is irreducible under the action of
$\mathfrak{g}' \cap \mathfrak{k}$. The generator $X_{A''0} \in \mathfrak{g}$
(for $A'' \in \{1, \ldots, n\})$ sends $W(k)$ to $W(k) \oplus W(k\pm 1)$ and
so does $[C', X_{A'' 0}]$, $C'$ being the Casimir of $\mathfrak{so}(n)$,
so that, for every $k\in \mathbb N$, there is a shift operator
$X_{A''}^\pm \in d({\mathcal U}')$, linear combination of $X_{A''0}$ and
$[C', X_{A'' 0}]$, sending $W(k)$ to $W(k\pm 1)$. Every $W(k)$ being of
multiplicity one, $d({\mathcal U}') \varphi$ contains every
$\mathfrak{k}$-type of $d$, so that the closure of $V_0$ is $\mathcal H$ and the
restriction to $G'$ is irreducible.

\md

If $G'={\mathcal S}^-_n$, the situation is somewhat more complicated.
 Let $e_1$ be the stabilized vector , so that $\mathfrak{x}$ is spanned by
$\{X_{1A}\}, \mathfrak{k}\cap \mathfrak{g}'$ being isomorphic to
$\mathfrak{so}(2) \oplus \mathfrak{so}(n-1)$.
Assume first $s=0$, so that $\mathcal H$ contains a trivial
$\mathfrak{so}(n)$-submodule $W(0)$. Let $\varphi \in W(0)$:
clearly $X_{1A'}\varphi  = 0$ if $A'\in \{2, \ldots, n\}$ and
$X\left(\scriptscriptstyle{\stackrel{-\varepsilon}{0}}\right)_1 \varphi = 0$
too, so that $\mathfrak{x}W(0)$ is spanned by
$X\left(\scriptscriptstyle{\stackrel{+\varepsilon}{0}}\right)_1 \varphi
= \varphi^+$. Since $\mathfrak{so}(n-1)$ commutes with
$X\left(\scriptscriptstyle{\stackrel{+ \varepsilon}{0}}
\right)_1$, it stills act
trivially on $\mathfrak{x}W(0)$, while the eigenvalue of $H_0$ increases by 1
in absolute value, so that $\mathfrak{k}\cap \mathfrak{g}'$ stabilizes
$\mathfrak{x} W(0)$.  Moreover, for $A' \in\{2, \ldots, n\}$,

\begin{equation}
X\left(\scriptscriptstyle{\stackrel{-\varepsilon}{0}}\right)_{A'} \varphi^+ =
(X\left(\scriptscriptstyle{\stackrel{\varepsilon}{0}}\right)_{1}
X\left(\scriptscriptstyle{\stackrel{-\varepsilon}{0}}\right)_{A'} -
[X\left(\scriptscriptstyle{\stackrel{\varepsilon}{0}}\right)_{1}, \hskip.2cm
X\left(\scriptscriptstyle{\stackrel{-\varepsilon}{0}}\right)_{A'}]) \varphi = 2
\varepsilon \sqrt{-1} \hskip.2cmX_{1A'} \varphi =0
\end{equation}

\noindent so $\varphi^+$ is an extremal weight vector of $\mathfrak{g}'$,
as well as $\varphi$, so that $V_0 \cap V_1 = \{0\}$.

Assume next $s \neq 0$ and $n$ even $(n \geq 4)$. Since $d(H_1) = \pm d(H_a)$
on an extremal vector for every $W(k)$, one has
$d(\sqrt{-1}\hskip.2cm X_{12}) \varphi = \pm s \varphi \neq
0$ on an extremal vector of $W(0)$, so that

\begin{equation}
{\mathcal U} (\mathfrak{so}(n-1)). (\mathfrak{x}\cap \mathfrak{k}) W(0) =
 {\mathcal U}(\mathfrak{so}(n-1)) W(0) ={\mathcal U}
(\mathfrak{so}(n)) W(0)=W(0)
\end{equation}

\noindent and $V_0 = V_1$.

Assume finally $n$ odd and $s = \frac{1}{2}$. The lowest
$\mathfrak{so}(n)$-type is a spinorial representation, and it is well known
that such a representation of $\mathfrak{so}(2r+1)(r\in \mathbb N)$ splits
into two inequivalent spinorial representations of $\mathfrak{so}(2r)$ of
equal dimensions; they are labelled
${\mathcal D}( \frac{1}{2}, \ldots, \frac{1}{2}, \pm \frac{1}{2})$ with the
two different choices of sign.

\md

Summarizing one has:

\md
\noindent {\bf{Proposition 3.1: }} The representation
$d^{n, \varepsilon}_{(-(s+\frac{n}{2}-1), \vec{s})}$ remains irreducible when
restricted  to $SO_0(1, n)$. Its restriction on
$SO_0(2, n - 1)$ when $s=0$ is the direct sum

$$d^{n-1, \varepsilon}_{(-(\frac{n}{2}-1),
\vec{0})} \oplus d^{n-1, \varepsilon}_{(-\frac{n}{2}, \vec{0})};$$

\noindent for $s=\frac{1}{2}$ and $n=2r+1$ odd, its restriction is

$$d^{2r, \varepsilon}_{(-r, \frac{1}{2}, \ldots, \frac{1}{2}, +\frac{1}{2})}
 \oplus d^{2r,\varepsilon}_{(-r, \frac{1}{2}, \ldots, \frac{1}{2},
 -\frac{1}{2})};$$

\noindent for $s\neq 0$ and $n=2r$
even, the restriction is irreducible and equal to

$$d^{2r-1,\varepsilon}_{(-(s+r-1),\partial \vec{s})},$$

\noindent where $\partial\vec{s}$ comes from $\vec{s}=(s,\ldots,s,\pm s)$
by dropping the last coordinate $\pm s$.
\hfill$\Box$

\bigskip

\noindent{\bf{c. Contraction of representations}}

\md

The Wigner-Inon\"{u} contraction of Lie algebras \cite{WinIn} can be defined
as follows: given a Lie algebra $\mathfrak{g}$ and a continuous family
$\Phi_\alpha \in GL(\mathfrak{g})$ of linear transformations of the underlying
vector space, with $0< \alpha \leq 1$ and $\Phi_1= 1$, a Lie algebra
$\mathfrak{g}_\alpha$ isomorphic to $\mathfrak{g}$ is defined on the same
underlying space by the Lie bracket:

\begin{equation}
[X, Y]_\alpha = \Phi^{-1}_{\alpha} [\Phi_\alpha X, \Phi_\alpha Y].
\end{equation}

If $\displaystyle\lim_{\alpha\longrightarrow 0}(\Phi_\alpha)$
is a non invertible mapping and $[X, Y]_0
= \lim([X, Y]_\alpha)$ exists when $\alpha \longrightarrow 0$, the Lie algebra
$\mathfrak{g}_0$ defined on the same underlying space is the contracted of
$\mathfrak{g}$ by the family $\{\Phi_\alpha\}$.

\medskip

Contraction of representations $U_\alpha$ of $\mathfrak{g}_\alpha$ on
${\mathcal H}_\alpha$ are defined in analogy. Here we shall limit ourselves to a
fixed representation space $\mathcal H$. Given a continuous family
$\{Z_\alpha\}$
of closed invertible linear transformations of $\mathcal H$ for
 $0<\alpha\leq 1$
with $Z_1=1$, and a representation $U_1 = U$ of $\mathfrak{g}_1 =
\mathfrak{g}$, defined on a dense domain $\mathcal E$ of analytic vectors,
the map

\begin{equation}
X\longmapsto U_\alpha (X) = Z_\alpha^{-1} U(\Phi_\alpha X) Z_\alpha
\end{equation}

\noindent is a representation of $\mathfrak{g}_\alpha$; indeed, one has:

\begin{equation}
\begin{array}{lcl}
[U_\alpha(X), U_\alpha (Y)] &=& Z^{-1}_\alpha [U (\Phi_\alpha Y),
U (\Phi_\alpha Y)] Z_\alpha\cr
  &  &  \cr
 &=&  Z^{-1}_\alpha U ([\Phi_\alpha X, \Phi_\alpha Y]) Z_\alpha\cr
    &   &  \cr
&=&  Z^{-1}_\alpha U (\Phi_\alpha [X, Y]_\alpha) Z_\alpha =
 U_\alpha ([X, Y]_\alpha) \end{array}\end{equation}

If the limit of $U_\alpha(X)$ exists for every $X\in\mathfrak{g}$ when
$\alpha\longrightarrow 0$ (regardless to whether $Z_\alpha$ has a limit),
then $\displaystyle U_0 =\lim_{\alpha\longrightarrow 0} U_\alpha$ is a
representation of the contracted Lie algebra $\mathfrak{g}_0$: we shall say
that it is the contracted of $U_1$ through the family $(Z_\alpha)$.

\medskip

Let us apply this to $\mathfrak{g}_1 =$ Lie
$({\mathcal S}^\pm_n)= \mathfrak{l}_n
\oplus \mathfrak{y}$  where $\mathfrak{y}$ is spanned by the generators
 $Y_\mu = X_{\mu w}, \mu \in J$, with $w=n$ for ${\mathcal S}^+_n$ and $w=-1$
 for ${\mathcal S}^-_n$; one has $[\mathfrak{y}, \mathfrak{y}] =
\mathfrak{l}_n$. We shall define the familly $\{\Phi_\alpha\}$ by:

\begin{equation}
\Phi_\alpha(X_{\mu \nu}) = X_{\mu \nu};\hskip.3cm \Phi_\alpha (Y_\mu) =
 \alpha(2-\alpha) Y_\mu.
\end{equation}

\md

Clearly, one has

\begin{equation}
[Y_\mu, Y_\nu]_\alpha = \alpha^2 (2-\alpha)^2[Y_\mu, Y_\nu];
\hskip.3cm [X_{\mu \nu}, Y_\lambda]_\alpha = [X_{\mu \nu}, Y_\lambda]
\end{equation}

\noindent so that the contacted algebra $\mathfrak{g}_0$ is isomorphic
to $\mathfrak{p}_n$.

\md

Let now $d$ be a massless representation of $\bar{G}_n$ on $\mathcal H$,
with analytic domain $\mathcal E$, on which all operators of the Lie algebra
are defined, their expressions being given by (1.20) (in particular $d(P_\mu)=
\sqrt{-1}\hskip.2cm x_\mu$) and (2.50).

Let $U=U_1$ be the restriction of $d$ to $\mathfrak{g}_1$, so that one has

\begin{equation}
U(Y_\mu) = \frac{1}{2}(d (P_\mu)\mp d(\hat{P}_\mu)).
\end{equation}

Define the family $\{Z_\alpha\}$ by

\begin{equation}
(Z_\alpha \varphi) (x) = \alpha^{(n-2)/2} \varphi (\alpha x)
\end{equation}

\noindent $Z_\alpha$ is a unitary operator, equal to
$\exp(d(\mathrm{Log}\alpha\, X_{n,-1}))$,
which has no limit for $\alpha \longrightarrow 0$. It satisfies:
\begin{equation}
Z^{-1}_\alpha d(P_\mu) Z_\alpha = \alpha^{-1}
d(P_\mu);\hskip.3cm Z^{-1}_\alpha d(\hat{P}_\mu)
Z_\alpha = \alpha\hskip.1cm d\hskip.1cm (\hat{P}_\mu)
\end{equation}

\noindent so that:

\begin{equation}
U_\alpha (Y_\mu) = (1 - \frac{\alpha}{2})(d(P_\mu) \mp \alpha^2
d(\hat{P}_\mu))
\end{equation}

\noindent while $U_\alpha(X_{\mu \nu}) = U(X_{\mu \nu})$. It is clear that
the limit of $U_\alpha(Y_\mu)$ exists for $\alpha \longrightarrow 0$, and it
is equal to $d(P_\mu)$. We have thus proved:

\md

\noindent {\bf{Proposition 3.2: }} The restriction $U$ on
$\overline{\mathcal S}^\pm_n$
of the massless representation $d$ of $\bar{G}_n$ contracts to its restriction
on $\overline{\mathcal P}_n$ through the family of unitary operators
$\{Z_\alpha\}$.
\hfill$\Box$

\newpage


\section{Conclusion}

Comparing the results obtained here with the classical case
$n=4$, we first observe that the main features are conserved:
only massless representations $U^S$ of $\overline{\mathcal P}_n$ can be
extended to ones of $\overline{G}_n$, and when this is possible the extension
$d$ is unique: it is a unitary irreducible representation with extremal weight,
vanishing on the two-sided ideal of the enveloping algebra generated by
$P_\mu P^\mu$. The form of the remaining Lie algebra generators is
completely determined when those of $\mathfrak{p}_n$ are given (that is,
$d$ is not only fixed up to equivalence, but when $U^S$ is fixed inside
its equivalence class so is $d$).

\medskip

Moreover, $d$ is a representation of either $G_n$ itself
(when $U^S$ is one of ${\mathcal P}_n$, that is $s$ integer) or of
a twofold covering (when $s$ is half-integer). All
representations $d$ are realizable on a functional space over the
corresponding Minkowski space or over a half-cone of its Fourier
dual (in fact, when $S$ is trivial $d$ is equivalent to the representation
induced by the trivial representation of the parabolic subgroup
${\mathcal W}_n$).

\medskip

The only feature which does not generalize concerns the restrictions
imposed on the inducing representation $S$. For $n=4$, the only
restriction is that $S$ is trivial on the translation subgroup $T_2$
of the twofold covering of the Euclidean group $E_2$.
This discards the so-called continuous spin representations and
allows all helicities $\pm s \in \frac{1}{2}\mathbb Z$.

\medskip

For  $n > 4$, $S$ must still vanish on the translations, but there are
additional constraints on $S$, depending on the parity of $n$: if $n= 2r$ is
even every coordinate of the extremal weight must equal in absolute value to
the last one (the minimal one), which is equal to $\pm s$. This constraint
is automatically satisfied for $n=4$, since $\mathfrak{so}(n-2)$ has rank 1
and the last coordinate of the weight is also the only one: the study of
the case $n=4$ alone gives no hint about this new constraint.

\medskip

For odd $n$ the constraints are quite drastic:
$S$ may be either trivial or spinorial. This appears as a straightforward
generalization of the case $n=3$ \cite{FFG}, if one defines
 $Spin (1)$ as $\mathbb Z_2$.

\medskip

If, instead of increasing $n$, one decreases it to $n=2$, one finds again
that the only UIR of the simply connected ${\mathcal P}_2 = SO_0(1, 1).T_2$
(besides the trivial one) which extend to
 $\overline{G}_2=\overline{SO}_0(2, 2)$ are the
massless ones: the massless orbits are the connected components of the
isotropic cone, that is, 4 half lines (instead of two half ones) and the
stabilizer is just $\{1\}$, so that massless UIR vanish on the subgroup
 spanned by $P_0 + P_1$ or $P_0 -P_1$, the factor group on which they are
faithful being here isomorphic to the connected affine group  $(x\longmapsto
ax+b)$ of the real line. By theorem 2.2, the extension to
$\mathfrak{so}(2,2) = \mathfrak{u}^+\oplus \mathfrak{u}^-$ must vanish on one
of the two factors $\mathfrak{u}^\pm$ (both isomorphic to
$\mathfrak{so}(2,1)$). The Casimir operator may take any value $C$ and by
(2.46) one gets $\hat{P}_0 = (P_0)^{-1}(D^2 - D+ \frac{1}{2} C)$.
There is no uniqueness of the extension, not even unitarity
($C$ may be any complex number):
lowering $n$ to 2 removes all constraints. One should however mention that
in this case the full conformal group is infinite, as is well-known.
We shall not discuss this case further here.

\medskip

Concerning the Poincar\'e - De Sitter relations, the sequence ``extension to
$\overline{G}_n$, then restriction to ${\mathcal S}^\pm_n$, then contraction to
$\overline{\mathcal P}_n$" is cyclic for every $n\geq 3$;  the demonstration is
practically identical with the one for $n=4$ \cite{AFFS}. As for the
irreduciblility of the restriction to the real rank two De Sitter subgroup
${\mathcal S}^-_n$, the result for $n$ even is a straightforward generalization
of the case $n=4$: the restriction splits into two simple factors if the
inducing representation is trivial, otherwise it is irreducible.
When $n$ is odd, it splits into two simple factors for both $s=0$ and
$s=\frac{1}{2}$.  The restriction on
$\overline{\mathcal S}^+_n = \overline{SO}_0(1, n)$ is always irreducible.

\medskip

Finally we should mention that there are some interesting open problems
involving massless representations, such as their tensor products with other
representations (in particular their tensor squares) or their appearance as
factors in indecomposable representations.

\newpage

\end{document}